\documentclass[a4paper,12pt]{article}
\usepackage{amsmath,enumerate}
\usepackage{graphicx}
\usepackage{multirow}
\usepackage{xcolor}
\definecolor{MyBlue}{rgb}{0.25,0.5,0.75}
\usepackage[colorlinks=true,allcolors=blue]{hyperref}
\newcommand{\lapprox}{%
	\mathrel{%
		\setbox0=\hbox{$<$}
		\raise0.6ex\copy0\kern-\wd0
		\lower0.65ex\hbox{$\sim$}
}}
\newcommand{\gapprox}{%
	\mathrel{%
		\setbox0=\hbox{$>$}
		\raise0.6ex\copy0\kern-\wd0
		\lower0.65ex\hbox{$\sim$}
}}
\begin{document}
	\begin{center}
		
		{\Large \bf Fermion mass hierarchy and lepton flavor violation using $CP$ symmetry}\\[20mm]
		Joy Ganguly \\
		Department of Physics\\
		Indian Institute of Technology Hyderabad,\\
		Kandi - 502 284, India.\\[3mm]
		Email: joyganguli2013@gmail.com\\[20mm]
		
	\end{center}
	\begin{abstract}
		We present a model which employs $CP\times Z_3$ symmetries where quark mixing and charged leptons masses are explained by following some texture. To achieve neutrino mass and mixing, we write a non-renormalizable Lagrangian using type-II seesaw mechanism by assuming ${\mathcal{O}}(1)$ couplings predicting no $CP$ violation. Then, we calculate the value of the couplings for both normal and inverted ordering of neutrino masses. We propose a mechanism of ultra-violet (UV) completion that motivates the higher dimensional Lagrangian in the neutrino sector. We also calculate the branching fraction of lepton flavor violating decays driven by triplet scalar for the model. Later, it is shown that the model can also accommodate maximal $CP$ violation by breaking the $CP$ symmetry spontaneously.
	\end{abstract}
	
	\newpage
	\section{Introduction}
	The Standard Model (SM) of particle physics is very successful in describing the
	observed phenomenology in different energy scales. However, it has limitations, like evidence of neutrino masses and hierarchy in Yukawa couplings.
	Neutrino oscillation data \cite{deSalas:2020pgw} suggests that neutrinos have a tiny mass and large mixing. In the SM, all the three neutrinos are within the $SU(2)$ doublets, and as a result, they are massless. Hence, one needs to go beyond the SM to generate
	neutrino masses. Many scenarios beyond the SM invoke tiny neutrino
	masses among which the seesaw mechanism \cite{Minkowski:1977sc} is the most promising scenario at the tree level. There are three types of canonical seesaw mechanisms depending on the new particles added to the SM.
	In type I \cite{Minkowski:1977sc,Mohapatra:1979ia},
	type II \cite{Magg:1980ut}, and type III \cite{Foot:1988aq} seesaw, right-handed neutrinos, $SU(2)$ triplet scalar, and $SU(2)$
	triplet fermions are exchanged, respectively. Many models have been proposed in explaining the lepton mixing, and neutrino oscillation data indicates that the lepton mixing is close
	to Tri-bimaximal mixing(TBM) \cite{Harrison:2002er}. In TBM pattern three lepton mixing angles have
	values: $\sin^2 \theta_{23}=\frac{1}{2}$, $\sin^2 \theta_{12}=\frac{1}{3}$ and $\sin^2 \theta_{13}=0$ though the best-fit value of $\sin \theta_{13}$ is approximately 0.15 \cite{deSalas:2020pgw}.
	So, to probe leptons mixing, new models based on $\mu-\tau$ symmetry \cite{Fukuyama:1997ky}, $\mu-\tau$ reflection symmetry \cite{Harrison:2002et}, and TBM mixing \cite{Harrison:2002er} have been built which extend the SM
	in different ways. There are still some unanswered questions to be found in experiments
	like the ordering of neutrino masses that is Normal Ordering (NO) or Inverted Ordering
	(IO) and the nature of neutrinos: Dirac or Majorana.
	
	When one explains lepton mixing in the neutrino sector, lepton flavor violating couplings appear in the model. As a result, lepton flavor violating (LFV) processes like $l\rightarrow l^{\prime}\gamma$, $l\rightarrow 3 l^{\prime}$, $l\rightarrow l^{\prime}$ occur. Here $l$, $l^{\prime}$ signifies different charged lepton families. However, till now, there is no experimental proof of these LFV processes, and as a result upper limit on the branching fraction of these processes has been calculated \cite{quark PDG}. These LFV processes can constrain the neutrino mass models in different ways which have been discussed in \cite{Masiero:2002jn} and references therein.
	
	In this work, we focused on explaining the quark mass hierarchy and mixing, charged lepton mass hierarchy, and neutrino mixing in a consistent scenario by using higher-dimensional terms in the $CP$ symmetric Lagrangian where all the couplings are of $\mathcal{O}(1)$. Previously Refs. \cite{Babu:1999me}, $\cite{Lykken:2008bw}$ explained quark mass and mixing and charged lepton masses using higher-dimensional terms in the Lagrangian, but they did not explain the neutrino sector within the same scenario. Though the mass matrix structure of quarks and charged leptons in our framework is similar to these mentioned references, the mass matrix texture has been motivated from different symmetry prospects. In addition to that, we incorporated the neutrino sector within the same framework and could explain all the neutrino oscillation data by diagonalizing the mass matrix \cite{Ganguly:2020riw}. In our approach, due to $CP$ symmetry, there is no CP-violating phases appearing in the model. But later, we showed that spontaneous breaking of $CP$ symmetry generates maximal $CP$ violation in the same model. To see some works where only neutrino masses and mixing have been explained by using higher-dimensional terms in the Lagrangian, please refer to \cite{Giudice:2008uua,Babu:2009aq,Bonnet:2009ej,Picek:2009is}. To see some works in the direction of quark mixing, see \cite{Ganguly:2021swa,rr}. However, our approach in this work is different. In Ref. \cite{Lykken:2008bw}, the higher dimensional terms in the Lagrangian are motivated by the UV completion mechanism. In our scenario, as texture of quarks and charged leptons mass matrices are the same to their case, we do not show it here explicitly. To the  best of our knowledge the texture of neutrino mass matrix has never been motivated in some UV complete Lagrangian, renormalizable at some larger scale, which we have explicitly discussed in the corresponding section of the paper. We have explicitly discussed the mechanism to make our scenario UV complete. We know that the type II seesaw mechanism drives lepton flavor violation. So, here we have studied the branching fraction of the LFV decays. Then, using the experimental constraints, we have shown the lower bound on the mass of the scalars, which facilitated the LFV decays. 
	
	The rest of the paper is organized in the following manner. In the next section, we have given symmetries and particles of our model. In sections \ref{sec:quark mass and mixing} and \ref{sec:charged leptons}, we evaluate the quark and charged lepton masses numerically. In section \ref{sec:neutrino mixing}, we calculate neutrino mass and mixing by writing a non-renormalizable Lagrangian, which can predict the NO of neutrino masses when the triplet vacuum expectation value is 1 eV, and it is shown that this Lagrangian is motivated from some UV complete Lagrangian at some high energy scale. In Section \ref{sec:LFV} we discussed the LFV decays in our model. In section \ref{sec:scalar potential}, we have given the full scalar potential. Then, in section \ref{sec:cp violation model}, we show how maximal $CP$ violation can be incorporated into the same framework and finally, we conclude in the section \ref{sec:conclusion}.
	\section{Our Model}\label{sec:our model}
	We are considering a model where we extend the SM by one more Higgs doublet and one Higgs triplet and two singlet fields $X$ and $X^{\prime}$. Three $SU(2)$ lepton doublets are denoted by $D_{\alpha L}=(\nu_{\alpha L}, \alpha_L)$ while the three right-handed singlets are denoted as $\alpha_R$, where $\alpha=e, \mu, \tau$ respectively. The quark sector also contains three quark doublets, up and down type quarks denoted as $Q_{\beta L}$, $u_{\beta R}$ and $d_{\beta R}$ respectively where $\beta$=1,2,3 respectively. The Higgs doublets and triplet are represented by
	\begin{eqnarray}\label{eq:scalar fields def}
	\phi_{i}=\begin{pmatrix}
	\phi_{i}^+ \\
	\phi_{i}^0
	\end{pmatrix},\hspace{4mm}
	\Delta=\begin{pmatrix}
	\frac{\Delta^+}{\sqrt{2}} & \Delta^{++} \\
	\Delta^0 & -\frac{\Delta}{\sqrt{2}}
	\end{pmatrix}, \hspace{4mm} i=1, 2.
	\end{eqnarray}
	Then, we introduce a CP symmetry defined as
	\begin{eqnarray}\label{CP transformation}
	&&	D_{\alpha L}\rightarrow i \gamma^0 C \bar{D}_{\alpha L}^T, \hspace{4mm} \alpha_R \rightarrow i \gamma^0 C \bar{\alpha}_R^T, \nonumber \\
	&& Q_{\beta L} \rightarrow i \gamma^0 C \bar{Q}_{\beta L}^T, \hspace{4mm} u_{\beta R} \rightarrow i \gamma^0 C \bar{u}_{\beta R}^T, \hspace{4mm}d_{\beta R} \rightarrow i \gamma^0 C \bar{d}_{\beta R}^T \nonumber  \\
	&& \phi_1 \rightarrow \phi_1^*, \hspace{4mm} \phi_2 \rightarrow -\phi_2^*, \hspace{4mm} \Delta \rightarrow \Delta^*\quad X\rightarrow X^{*},\quad X^{\prime}\rightarrow {X^{\prime}}^*.
	\end{eqnarray}
	where C is the charge conjugation matrix. In addition to the above CP symmetry, we will impose one $Z_3$ symmetry. We will define the charge assignment of the fields mentioned above under $Z_3$ in the respective sections with the motivation. Here, we discuss the role of the scalar fields in our model. Two Higgs doublets are instrumental in generating the masses of quarks and charged leptons, while the triplet is responsible for neutrino mass generation. The singlet fields $X$ is liable to cause the hierarchical structure of the quark mass matrix and charged lepton mass matrix, which we will discuss in the next section. Here it is worth noticing that Both $X$ and $X^{\prime}$ are needed to generate the correct hierarchy in neutrino masses.
	
	The neutral components of doublet and triplet scalar fields acquire non-zero vacuum expectation values (VEVs) $\langle 0|\phi_j^0|0 \rangle = v_{j}/\sqrt{2}$ and $\langle 0|\Delta^0|0 \rangle = v_{\Delta}$, where $\sqrt{v_1^2+v_2^2+2v_{\Delta}^2}=246 \hspace{1mm}\rm{GeV}$ that represents the electroweak scale. Now, CP conservation requires $v_1^*=v_1$ and $v_2^*=-v_2$. But in our model, to break the CP symmetry spontaneously, both the VEVs are taken to be real. This CP violation is necessary in order to explain the CP violating phase in the quark sector. The triplet VEV is constrained by the $\rho$ parameter which involves the ratio of W and Z boson masses. The SM value of the $\rho$ parameter is 1 which is in perfect agreement with the eletroweak precision measurements $\rho_{\rm{obs}}=1.00039 \pm 0.00019$ \cite{quark PDG} and this constrains the VEV of triplet $v_{\Delta}$ to be less than 2 GeV \cite{rho param}. 
	
	\section {Quark Mass and Mixing}\label{sec:quark mass and mixing}
	In this model, two Higgs doublets will give masses to quarks with our model's $CP$ and other symmetries. We know that quark masses are hierarchical. One can explain the quark mixing if the Yukawa couplings are hierarchically suppressed. There is one model proposed by Babu and Nandi \cite{Babu:1999me} where they explain quark mixing patterns through hierarchically suppressed Yukawa couplings. Later this model is modified in Ref. \cite{Lykken:2008bw} where the suppression in Yukawa couplings is explained with a singlet scalar field $X$. We follow the similar mechanism of \cite{Lykken:2008bw} in our framework to illustrate the quark mixing pattern.
	
	In our model, all the quark fields and scalar doublets are assumed to be singlet under $Z_3$ while the $X$ field transform under $Z_3$ as $X\rightarrow \omega X$. Now we can write $CP\times Z_3$ invariant Lagrangian for quarks in our model with the fields and symmetries mentioned above. To write the Yukawa couplings, for simplifying the notation, we designate the following quantities,
	\begin{eqnarray}\label{h_jk define}
	h_{jk}^u \tilde{\phi} \equiv h_{jk}^{u^{\prime}} \tilde{\phi}_1+ i h_{jk}^{u^{\prime \prime}} \tilde{\phi}_2, \hspace{4mm} h_{jk}^d \phi \equiv h_{jk}^{d^{\prime}} \phi_1+ i h_{jk}^{d^{\prime \prime}} \phi_2,
	\end{eqnarray}
	Here $h_{jk}^{u^{\prime}}, h_{jk}^{u^{\prime \prime}}$, $h_{jk}^{d^{\prime}}$, $h_{jk}^{d^{\prime \prime}}$ are dimensionless quantities.
	
	The following Lagrangian for quark Yukawa couplings can be written in our model 
	 as
	\begin{eqnarray}\label{quark Lag}
	\mathcal{L}_Y&=&\bar{Q}_{3L}h_{33}^u \tilde{\phi}u_{3R}+ \frac{X^{\dagger}X}{M^2}[\bar{Q}_{3L} h_{33}^d\phi d_{3R}+\bar{Q}_{2L}h_{22}^u \tilde{\phi}u_{2R}+\bar{Q}_{2L}h_{23}^u \tilde{\phi}u_{3R}+ \bar{Q}_{3L}h_{32}^u \tilde{\phi}u_{2R}] \nonumber \\
	&& \Big(\frac{X^{\dagger}X}{M^2}\Big)^2[\bar{Q}_{2L}h_{22}^d \phi d_{2R}+\bar{Q}_{2L}h_{23}^d \phi d_{3R} + \bar{Q}_{3L}h_{32}^d \phi d_{2R} + \bar{Q}_{1L}h_{12}^u \tilde{\phi}u_{2R} \nonumber \\
	&& + \bar{Q}_{2L}h_{21}^u \tilde{\phi}u_{1R}+ \bar{Q}_{1L}h_{13}^u \tilde{\phi}u_{3R} + \bar{Q}_{3L}h_{31}^u \tilde{\phi}u_{1R}] + 
\Big(\frac{X^{\dagger}X}{M^2}\Big)^3[\bar{Q}_{1L}h_{11}^u \tilde{\phi}u_{1R} \nonumber \\
	&&+ \bar{Q}_{1L}h_{11}^d \phi d_{1R} + \bar{Q}_{1L}h_{12}^d \phi d_{2R} + \bar{Q}_{2L}h_{21}^d \phi d_{1R} + \bar{Q}_{1L}h_{13}^d \phi d_{3R} + \bar{Q}_{3L}h_{31}^d \phi d_{1R}].
	\end{eqnarray}
	Here $M$ is the cut-off scale of our model, which can be around 1 TeV \cite{Babu:1999me,Lykken:2008bw}. In the above equation, the term $\frac{X^{\dagger}X}{M^2}$ gives suppression to quark Yukawa couplings. The higher dimensional terms in Eq. (\ref{quark Lag}) can be motivated through the study of UV completion for this model. Some additional symmetries and vector-like quarks have been proposed in the UV completion, which gains mass around the scale $M$ \cite{Lykken:2008bw}. After integrating out the heavy vector-like quarks, the non-renormalizable terms in Eq. (\ref {quark Lag}) appear below scale $M$. We will give a detailed description later. The Yukawa couplings $h_{jk}^{u^{\prime}}$, $h_{jk}^{u^{\prime \prime}}$, $h_{jk}^{d^{\prime}}$ and $h_{jk}^{d^{\prime \prime}}$ are real due to CP symmetry of Eq. (\ref{CP transformation}).
	
	After electroweak symmetry breaking, when both $\phi_{1,2}$ get VEVs, mass terms for quarks will be generated. As the triplet VEV is small, then we can take $v_1=v_2=v/\sqrt{2}$. Thus, Eq. (\ref{h_jk define}) takes a form,
	\begin{eqnarray}\label{h_u,d tilde}
	&&	\langle h_{jk}^{u} \tilde{\phi} \rangle =\tilde{h}_{jk}^{u}\begin{pmatrix}
	\frac{v}{2} \\
	0
	\end{pmatrix}, \tilde{h}_{jk}^u =h_{jk}^{u^{\prime}}+i h_{jk}^{u^{\prime \prime}},\nonumber \\
	&&	\langle h_{jk}^{d} \phi \rangle =\tilde{h}_{jk}^{d}\begin{pmatrix}
	\frac{v}{2} \\
	0
	\end{pmatrix}, \tilde{h}_{jk}^d =h_{jk}^{d^{\prime}}+i h_{jk}^{d^{\prime \prime}}. 	
	\end{eqnarray}
	Using the above relations into Eq. (\ref{quark Lag}), we can obtain the masses for up and down type quarks respectively, as
	\begin{eqnarray}\label{quark mass matrix}
	M_u=\begin{pmatrix}
	\tilde{h}_{11}^{u} \epsilon^6 & 	\tilde{h}_{12}^{u} \epsilon^4 & 	\tilde{h}_{13}^{u} \epsilon^4 \\
	\tilde{h}_{21}^{u} \epsilon^4 & 	\tilde{h}_{22}^{u} \epsilon^2 & 	\tilde{h}_{23}^{u} \epsilon^2 \\
	\tilde{h}_{31}^{u} \epsilon^4 & 	\tilde{h}_{32}^{u} \epsilon^2 & 	\tilde{h}_{33}^{u} 
	\end{pmatrix} \frac{v}{2},\hspace{4mm}
	M_d =\begin{pmatrix}
	\tilde{h}_{11}^{d} \epsilon^6 & 	\tilde{h}_{12}^{d} \epsilon^6 & 	\tilde{h}_{13}^{d} \epsilon^6 \\
	\tilde{h}_{21}^{d} \epsilon^6 & 	\tilde{h}_{22}^{d} \epsilon^4 & 	\tilde{h}_{23}^{d} \epsilon^4 \\
	\tilde{h}_{31}^{d} \epsilon^6 & 	\tilde{h}_{32}^{d} \epsilon^4 & 	\tilde{h}_{33}^{d}\epsilon^2 
	\end{pmatrix}\frac{v}{2}.
	\end{eqnarray}
	Here $\epsilon^2=\frac{\langle X^{\dagger}X \rangle}{M^2}=\frac{v_X^2}{M}$, where $\langle X \rangle =v_X$ can be complex in general. In the above equation, we can see that $\tilde{h}_{jk}^u $, $\tilde{h}_{jk}^d$ are complex, but as $\epsilon$ appears with even powers, these become real. Hence the matrices in Eq. (\ref {quark mass matrix}) are complex. Complex mass matrices for both up and down type quarks are required to explain CP violation. One thing to notice here is that instead of using $\phi_{1,2}$, the Lagrangian in Eq. (\ref{quark Lag}) can be made invariant either with $\phi_1$ or $\phi_2$. However, in the first case, if we use only $\phi_1$, all the Yukawa couplings will be real, and then one cannot explain the $CP$ violation. Similarly, using $\phi_2$ only makes all the Yukawa couplings imaginary. Then, up to an overall phase mass matrix for up and down type quarks will be real again, and explaining CP violation would be impossible. Also, using the two Higgs doublets make our model different from the Ref. \cite{Lykken:2008bw} where only one $\phi$ is used.
	
	After diagonalizing the mass matrices of Eq. (\ref{quark mass matrix}), the masses and mixing angles for quarks, up to leading order in $\epsilon$ can be written as \cite{Lykken:2008bw}
	\begin{eqnarray}\label{quark mixing mass}
	&&	(m_t,m_c,m_u) \approx (|\tilde{h}_{33}^u|, |\tilde{h}_{22}^u| \epsilon^2, |\tilde{h}_{11}^u-\tilde{h}_{12}^u \tilde{h}_{21}^u/\tilde{h}_{22}^u|\epsilon^6) v/2,\nonumber \\
	&& (m_b, m_s, m_d) \approx (|\tilde{h}_{33}^d|\epsilon^2, |\tilde{h}_{22}^d|\epsilon^4, |\tilde{h}_{11}^d|\epsilon^6)v/2,\nonumber \\
	&&|V_{us}| \approx \Bigg|\frac{\tilde{h}_{12}^d}{\tilde{h}_{22}^d}- \frac{\tilde{h}_{12}^u}{\tilde{h}_{22}^u}\Bigg| \epsilon^2,\nonumber \\
	&& |V_{cb}| \approx \Bigg|\frac{\tilde{h}_{23}^d}{\tilde{h}_{33}^d}- \frac{\tilde{h}_{23}^u}{\tilde{h}_{33}^u}\Bigg| \epsilon^2, \nonumber \\
	&& |V_{ub}| \approx \Bigg|\frac{\tilde{h}_{13}^d}{\tilde{h}_{33}^d}- \frac{\tilde{h}_{12}^u \tilde{h}_{23}^d}{\tilde{h}_{22}^u \tilde{h}_{33}^d}-\frac{\tilde{h}_{13}^u}{\tilde{h}_{33}^u}\Bigg| \epsilon^4,
	\end{eqnarray}
	We have fitted the above relations to the best fit values \cite{quark PDG},
	\begin{eqnarray}\label{PDG quark}
	&&	(m_t, m_c, m_u)=(172.76, 1.27, 2.16 \times 10^{-3})\hspace{1mm} \rm{GeV}, \nonumber \\
	&& (m_b, m_s, m_d)=(4.18 \times 10^3, 93, 4.67 )\hspace{1mm} \rm{MeV}, \nonumber \\
	&& (|V_{us}|, |V_{cb}|, |V_{ub}|) =(0.2245, 0.041, 0.00382).
	\end{eqnarray} 
	Now with $\epsilon=1/6$, the above mentioned fitting can be done with the following value of the Yukawa couplings,
	\begin{eqnarray}
	&& (|\tilde{h}_{33}^u|, |\tilde{h}_{22}^u| , |\tilde{h}_{11}^u-\tilde{h}_{12}^u \tilde{h}_{21}^u/\tilde{h}_{22}^u|)\approx (1.40, 0.37, 0.82),\nonumber \\
	&& (|\tilde{h}_{33}^d|, |\tilde{h}_{22}^d|, |\tilde{h}_{11}^d|) \approx	(1.22, 0.98, 1.77),\nonumber \\
	&& (|\tilde{h}_{12}^d|, |\tilde{h}_{12}^u|, |\tilde{h}_{23}^d|, |\tilde{h}_{23}^u|, |\tilde{h}_{13}^d|, |\tilde{h}_{13}^u| ) \approx (2.07, 2.2, 0.8, 1.15, 0.7, 0.7).
	\end{eqnarray} 
	It is worth mentioning that, with any $\epsilon$, smaller than $1/6$ some of the couplings get enhanced to be greater than 2.
	
	Hence with $\mathcal{O}(1)$ Yukawa couplings and $\epsilon=1/6$, we can explain the masses and mixing pattern of quarks. Since we are looking for the new physics to appear around 1 TeV, we can take the cut-off scale $M$ $\sim$ $1$ TeV and get $\langle X \rangle =167$ GeV. As elucidated before, the suppression in Yukawa couplings of Eq. (\ref {quark Lag}) is created by the factor $\frac{X^{\dagger}X}{M^2}$. Instead of $\frac{X^{\dagger}X}{M^2}$, the suppression can also be produced by $\frac{\phi_1^{\dagger}\phi_1}{M^2}$ or $\frac{\phi_2^{\dagger}\phi_2}{M^2}$ as has been done in Ref. \cite{Babu:1999me}. However in the latter case $\epsilon=v_1/2, v_2/2$. Since in our model $\epsilon=1/6$ can give good fitting to the quark mass and mixing pattern, then with the value of $v=246$ GeV, we will have $\langle X \rangle = 676$ GeV. It means new physics, which generates the higher dimensional terms of Eq. (\ref{quark Lag}), should appear around 676 GeV. Since there is no new physics found around 676 GeV in LHC, we cannot use $\frac{\phi_1^{\dagger}\phi_1}{M^2}$ or $\frac{\phi_2^{\dagger}\phi_2}{M^2}$ in place of $\frac{X^{\dagger}X}{M^2}$. This is why the singlet field $X$ is used to describe the suppression in quark Yukawa couplings.
	
	To interpret the origin of higher-dimensional terms in Eq.(\ref {quark Lag}), we follow the UV completion process described in \cite{Lykken:2008bw}. According to this process, additional symmetry $U(1)_X \times U(1)_F$ and singlet flavon fields F are introduced in addition to vector-like quarks. Both $X$ and $F$ fields are charged under $U(1)_X$ and $U(1)_F$ respectively, while the Higgs doublets are chosen to be singlet under $U(1)_X \times U(1)_F$. The flavor symmetry is spontaneously broken when $F$ acquires VEV around $M$. The charge assignment of all the fields is such that all the dimension 4 Yukawa coupling terms are forbidden except for the top quark. Now one can write invariant renormalizable terms between standard model quark fields and the extra fields. Upon integrating out the heavy vector-like quarks, below scale $M$, Eq. (\ref{quark Lag}) can be generated. The above-stated UV completion can be incorporated into our model, but then we have to impose the $CP \times Z_3$ on vector-like quarks and flavon fields. We have seen that this is straightforward as it is similar to $\cite{Lykken:2008bw}$. The scalar fields $X$ and $F$ can allow mixing between doublets and triplet Higgses of our model. We can assign zero charges for the triplet and doublets under $U(1)_X \times U(1)_F$. Then $X$ and $F$ fields can only enter into our model in the form of $X^{\dagger}X$ or $F^{\dagger}F$, and it will not affect the other VEVs.
	\section{Charged Lepton Mass}\label{sec:charged leptons}
	We will follow the same Ref. \cite{Lykken:2008bw} to explain charged lepton mass hierarchy. In this case replace $Q_{\beta L}$ by $D_{\alpha L}$, $d_{\beta R}$ by $\alpha_{R}$ and $h_{jk}^d$ by $h_{jk}^l$, then the mass matrix structure will be same as $M_d$ in Eq. (\ref{quark mass matrix}).
	\begin{eqnarray}\label{lepton mass matrix}
	M_l=\begin{pmatrix}
	\tilde{h}_{11}^l \epsilon^6 & \tilde{h}_{12}^l \epsilon^6 & \tilde{h}_{13}^l\epsilon^6 \\
	\tilde{h}_{21}^l \epsilon^6 & \tilde{h}_{22}^l \epsilon^4 & \tilde{h}_{23} \epsilon^4 \\
	\tilde{h}_{31}\epsilon^6 & \tilde{h}_{32}\epsilon^4 & \tilde{h}_{33}\epsilon^2 
	\end{pmatrix}\frac{v}{2}.
	\end{eqnarray}
	Here $\tilde{h}_{jk}^l$ has the same form as $\tilde{h}_{jk}^d$ in Eq. (\ref{h_u,d tilde}). $\epsilon^2$ and $v$ are already defined in the previous section.
	
	After diagonalizing the above mass matrix, up to leading order in $\epsilon$, the mass eigenvalues for three charged leptons are 
	\begin{eqnarray}\label{lepton eigenvalue}
	(m_{\tau}, m_{\mu}, m_e) \approx (|\tilde{h}_{33}^l| \epsilon^2, |\tilde{h}_{22}^l |\epsilon^4, |\tilde{h}_{11}^l| \epsilon^6) v/2.
	\end{eqnarray}
	We can fix  $h_{33}^l$, $h_{22}^l$, $h_{11}^l$ by appropriately fitting the above relation to the best fit values \cite{quark PDG},
	\begin{eqnarray}\label{lepton mass}
	(m_{\tau}, m_{\mu}, m_e)=(1.776, 0.1056, 5.11 \times 10^{-4}) \vspace{2mm} \rm{GeV}.
	\end{eqnarray} 
	Previously the quark mass and mixing have been fitted with the value of $\epsilon=1/6$; then, the same $\epsilon$ value should fix the Yukawa couplings as
	\begin{eqnarray}
	(|\tilde{h}_{33}^l|, |\tilde{h}_{22}^l|, |\tilde{h}_{11}^l|)\approx (0.52, 1.11, 0.20).
	\end{eqnarray}
	Hence the hierarchy of charged lepton mass matrix can be illustrated with the same $\epsilon$, which explains the quark mixing pattern. Here the numerical results follow the $\mathcal{O}(1)$ approximation quite well.
	
	The mass matrix in Eq. (\ref{lepton mass matrix}) can be diagonalized as follows,
	\begin{eqnarray}
	{V_{L}^l}^{\dagger} M_l M_l^{\dagger} V_{L}^l={\rm Diag}(m_e^2,m_{\mu}^2, m_{\tau}^2).
	\end{eqnarray}
	Here $V_{L}^l$ is the unitary  matrix which diagonalize the charged lepton mass matrix in Eq. (\ref{lepton mass matrix}). The form of $V_{L}^l$ can be found as \cite{Ghosh:2011up}
	\begin{eqnarray}
	V_{L}^l=\begin{pmatrix}
	1 & \frac{\tilde{h}_{12}^l}{\tilde{h}_{22}^l}\epsilon^2 & \frac{\tilde{h}_{13}^l}{\tilde{h}_{33}^l}\epsilon^4 \\
	-\frac{\tilde{h}_{12}^l}{\tilde{h}_{22}^l}\epsilon^2 & 1 & \frac{\tilde{h}_{23}^l}{\tilde{h}_{33}^l}\epsilon^2 \\
	-\frac{\tilde{h}_{13}^l \tilde{h}_{22}^l-\tilde{h}_{23}^l \tilde{h}_{12}^l}{\tilde{h}_{22}^l \tilde{h}_{33}^l} \epsilon^4 & \frac{\tilde{h}_{23}^l}{\tilde{h}_{33}^l}\epsilon^2 & 1 
	\end{pmatrix}.
	\end{eqnarray}
	The above form of $V_{L}^l$ suggests that it is close to the unit matrix with the off-diagonal elements suppressed by at least $\epsilon^2$. Hence we can say that the lepton mixing will be entirely determined by the neutrino sector, which we will show in the next section.
	
	The origin of the higher dimensional terms can be generated through the UV completion, as explained in the previous section for quarks. Here the charged lepton Lagrangian is similar to the down quark sector. Hence, the same UV completion is applicable for charged leptons with the suitable charge assignment under $U(1)_X \times U(1)_F$ with the introduction of heavy vector-like leptons and flavon fields. The charge assignment under $CP \times Z_3$ can be done accordingly.
	\section{Neutrino mass and mixing}\label{sec:neutrino mixing}
	In this section, we obtain neutrino mass and mixing pattern by writing non-renormalizable terms in the Lagrangian using type-II seesaw mechanism, which involves one $SU(2)_L$ triplet scalar $\Delta$ having hypercharge $Y=2$ which has the form given in Eq. (\ref {eq:scalar fields def}).
	\subsection{Model for neutrino mass and mixing}
	In this model, one triplet $\Delta$ is introduced to explain neutrino mass and mixing via type-II seesaw mechanism. The $CP$ symmetry of the left-handed lepton doublets and $\Delta$ are already defined in Eq. (\ref{CP transformation}). As we mentioned in the beginning that one additional scalar singlet $X^{\prime}$ is needed along with the $X$ field here to explain neutrino mass and mixing. Except for $X$, all the fields change trivially under $Z_3$.
	
	Neutrino oscillation data suggests that the mass scales for solar and atmospheric neutrinos are approximately 0.0087 and 0.05 eV \cite{deSalas:2020pgw}. So, for neutrino masses, the hierarchy is of the order of $\sim 10 $, but for the electron and tau, the mass hierarchy is of the order of $\sim 10^4$. As a result, we do not need an extensive hierarchy in the Yukawa couplings in the neutrino sector compared to the hierarchy in quark or charged leptons Yukawa couplings. 
	
	Hence, looking at the charge assignments and the naive order of estimation of neutrino mass, we propose the following higher-dimensional $CP \times Z_3$ Lagrangian,
	\begin{eqnarray}\label{neutrino lagrangian}
	&&	\mathcal{L}_Y =\Big(\frac{X}{M}\Big)^3 [Y_{ee}^{\nu} \bar{D}^c_{e L} i \sigma_2 \Delta D_{e L} + Y_{e\mu}^{\nu} \bar{D}^c_{e L} i \sigma_2 \Delta D_{\mu L} + Y_{e \tau}^{\nu} \bar{D}^c_{e L} i \sigma_2 \Delta D_{\tau L} ] \nonumber \\
	&& \Big(\frac{X^{\prime}}{M}\Big)^2 [Y_{\mu \mu}^{\nu} \bar{D}^c_{\mu L} i \sigma_2 \Delta D_{\mu L} +Y_{\tau \tau}^{\nu} \bar{D}^c_{\tau L} i \sigma_2 \Delta D_{\tau L} + Y_{\mu \tau}^{\nu} \bar{D}^c_{\mu L} i \sigma_2 \Delta D_{\tau L}] + h.c..
	\end{eqnarray}
	Here $\bar{D}^c_{\alpha L}$ is the charge conjugated doublet for $D_{\alpha L}$, where $\alpha =e, \mu, \tau $, and $\sigma_2$ is the Pauli matrix. Here it is assumed that all the couplings $Y_{\alpha \beta}^{\nu}$ to be of $\mathcal{O}(1)$, where $\alpha, \beta =e, \mu, \tau$. Application of CP symmetry defined in Eq. (\ref{CP transformation}) in the above equation makes all the $Y_{\alpha \beta}^{\nu}$ real. 
	
	The use of $Z_3$ symmetry on $X$ field can be recognized from the second line of the Lagrangian of Eq. (\ref{neutrino lagrangian}) where $X^2$ in place of ${X^{\prime}}^2$ cannot be written in an invariant way. Also, there are several higher dimensional terms possible like $(X/M)^2 \bar{D}_{eL}^ci\sigma_2\Delta D_{eL}$. However, when we later show the full UV completion of our model, one can understand that other possible higher dimensional terms cannot be generated. Before going into that, we will write the mass matrix and show the diagonalization such that one obtains $\mathcal{O}(1)$ couplings, by satisfying the neutrino oscillation data.
	
	The mass matrix can be written as
	\begin{eqnarray}\label{neutrino mass matrix}
	M_{\nu}=\begin{pmatrix}
	Y_{ee}^{\nu} \epsilon^3 &  Y_{e\mu}^{\nu}  \epsilon^3 & Y_{e \tau}^{\nu} \epsilon^3 \\
	Y_{e\mu}^{\nu}\epsilon^3 & Y_{\mu \mu}^{\nu}\delta^2 & Y_{\mu \tau}^{\nu}\delta^2 \\
	Y_{e\tau}^{\nu}\epsilon^3 & Y_{\mu \tau}^{\nu}\delta^2 & Y_{\tau \tau}^{\nu}\delta^2
	\end{pmatrix}v_{\Delta}.
	\end{eqnarray}
	Here $\delta=\frac{\langle  X^{\prime}\rangle}{M}$ and $\epsilon=\frac{\langle X \rangle}{M}$ both are real quantities.
	This mass matrix is symmetric. So, it can be diagonalized by a unitary matrix $V_{L}^{\nu}$ as
	\begin{eqnarray}\label{diagonalization 1}
	{V_L^{\nu}}^T M_{\nu} V_L^{\nu}={\rm{diag}}(m_1, m_2, m_3).
	\end{eqnarray}
	Here $m_1$, $m_2$, $m_3$ are three light neutrino masses.
	
	In the neutrino sector, our main goal is to fit the atmospheric and solar neutrino mass squared differences with $\mathcal{O}(1)$ couplings $Y^{\nu}$. The PMNS matrix \cite{Maki:1962mu,Pontecorvo:1967fh} is defined as $U_{PMNS}={V_L^l}^{\dagger}V_L^{\nu}$ where $V_L^l$ is the matrix which diagonalizes the charged lepton mass matrix. Previously in section \ref{sec:charged leptons}, it has been discussed that $V_L^l$ is almost diagonal. So, the lepton mixing pattern will be completely determined by the neutrino sector and the mixing matrix $V_L^{\nu}$.
	\subsection{Diagonalization}\label{subsec:diagonalization}
	$U_{\rm{PMNS}}$ is Pontecorvo-Maki-Nakagawa-Sakata(PMNS) \cite{Maki:1962mu,Pontecorvo:1967fh} matrix whose parametrization has been specified in terms of three lepton mixing angles and one Dirac CP violating phase according to the convention of PDG \cite{quark PDG}
	\begin{eqnarray}\label{PDG UPMNS}
	U_{\rm{PMNS}}=\begin{pmatrix}
	c_{12}c_{13} & s_{12}s_{13} & s_{13}e^{-i \delta_{\rm{CP}}} \\
	-s_{12}c_{23}-c_{12}s_{23}s_{13}e^{i\delta_{\rm{CP}}} & c_{12}c_{23}-s_{12}s_{23}s_{13}e^{i\delta_{\rm{CP}}} & s_{23}c_{13} \\
	s_{12}s_{23}-c_{12}c_{23}s_{13}e^{i\delta_{\rm{CP}}} & -c_{12}s_{23}-s_{12}c_{23}s_{13}e^{i\delta_{\rm{CP}}} &c_{23}c_{13}
	\end{pmatrix}.
	\end{eqnarray}
	Here $c_{ij}=\cos \theta_{ij}$, $s_{ij}=\sin \theta_{ij}$ and $\delta_{\rm{CP}}$ is the Dirac CP violating phase. If neutrinos are Majorana then there will be one diagonal phase matrix should be multiplied consists of two independent phases but as there are still no experimental evidences of that, we are assuming it to be zero for simplicity. Different neutrino oscillation data is well satisfied by the tri-bimaximal values \cite{Harrison:2002er} of the atmospheric and solar mixing angles $\theta_{23}$ and $\theta_{12}$ respectively which are designated as $\sin^2 \theta_{23}=\frac{1}{2}$ and $\sin^2 \theta_{12}=\frac{1}{3}$ though the TBM value of the reactor mixing angle ($\theta_{13}$) is ruled out. To be consistent with the experimental data we take $\delta_{\rm{CP}}=\pi$ which is within the $3 \sigma$ range of the experimental value \cite{deSalas:2020pgw}. So, putting $s_{23}^2=\frac{1}{2}$, $s_{12}^2=\frac{1}{3}$, $\delta_{\rm{CP}}=\pi$ in the $U_{\rm{PMNS}}$ in Eq. (\ref{PDG UPMNS}), we will have $V_{L}^{\nu}$ as
	\begin{eqnarray}\label{V l nu}
	V_L^{\nu}=\begin{pmatrix}
	\sqrt{\frac{2}{3}}c_{13} & \frac{c_{13}}{\sqrt{3}} & -s_{13} \\
	-\frac{1}{\sqrt{6}}+\frac{s_{13}}{\sqrt{3}} & \frac{1}{\sqrt{3}}+ \frac{s_{13}}{\sqrt{6}} & \frac{c_{13}}{\sqrt{2}} \\
	\frac{1}{\sqrt{6}}+\frac{s_{13}}{\sqrt{3}} & -\frac{1}{\sqrt{3}}+ \frac{s_{13}}{\sqrt{6}} & \frac{c_{13}}{\sqrt{2}}
	\end{pmatrix}.
	\end{eqnarray}
	
	Now the relation for diagonalizing the mass matrix in Eq. (\ref{diagonalization 1}) can be inverted and can be written as
	\begin{eqnarray}\label{inverted diagonalization formula}
	M_{\nu}=({V_L^{\nu}})^* diag(m_1, m_2, m_3) {V_L^{\nu}}^{\dagger}.
	\end{eqnarray}
	Here $s_{13}$ is the only small variable in $V_L^{\nu}$ in Eq. (\ref{V l nu}) of which the best-fit value is 0.15 \cite{deSalas:2020pgw}. Using the neutrino oscillation data, the following mass squared differences have been found \cite{deSalas:2020pgw}, the best-fit values are
	\begin{eqnarray}\label{neutrino oscc data}
	m_s^2 =m_2^2-m_1^2 =7.5 \times 10^{-5} eV^2,\hspace{3mm}m_a^2 = \begin{cases} m_3^2-m_1^2 =2.55 \times 10^{-3}& (\rm{NO}) \\ m_1^2-m_3^2=2.45 \times 10^{-3} & (\rm{IO}) \end{cases}.
	\end{eqnarray}
	Here $m_s^2$ and $m_a^2$ represent solar and atmospheric mass squared differences. $m_1$, $m_2$ and $m_3$ are the three light neutrino mass eigenvalues. Above neutrino oscillation data shows that $\frac{m_s^2}{m_a^2}\sim s_{13}^2$ can be neglected compared to the other mixing angle values. NO(IO) is meant for normal(inverted) hierarchy of neutrino masses. The above values give the mass scale of solar and atmospheric mass as $m_s \sim 0.0087$eV and $m_a\sim 0.05 $eV. The neutrino mass-squared differences can be fitted by the following neutrino masses
	\begin{eqnarray}
	{\rm{NO}}:\hspace{4mm} m_1 \leq m_s, \hspace{4mm} m_2 = \sqrt{m_s^2+m_1^2}, \hspace{4mm} m_3 =\sqrt{m_a^2+m_1^2}.\\
	{\rm{IO}}:\hspace{4mm} m_3 \leq m_s, \hspace{4mm} m_1 = \sqrt{m_a^2+m_3^2}, \hspace{4mm} m_2 =\sqrt{m_s^2+m_1^2}.
	\end{eqnarray}
	We can note from the above equation that $\frac{m_1}{m_a} \leq s_{13} $ in the case of NO but $\frac{m_1}{m_a} \leq 1 $ for IO. Similar approximations can be concluded for $\frac{m_2}{m_a}$ and $\frac{m_3}{m_a}$. We can find the expression for $\frac{1}{m_a}M_{\nu}$ up to first order in $\frac{m_s}{m_a}$, $s_{13}$ as
	\begin{eqnarray}\label{eq:expansion_mnu/ma}
	&&	\frac{1}{m_a}M_{\nu}=\frac{1}{m_a}\begin{pmatrix}
	p & q & r\\
	q & s & b \\
	r & b & s
	\end{pmatrix},\nonumber \\
	{\rm{NO}} &:& p=\frac{1}{3}(2 m_1+m_2),\hspace{3mm}q=\frac{1}{6}\{-3\sqrt{2}m_3 s_{13}-2 m_1+2 m_2\},\nonumber \\
	&&	r=\frac{1}{6}\{-3\sqrt{2}m_3 s_{13}+2 m_1-2 m_2\},\hspace{3mm} s=\frac{1}{6}\{3 m_3+2 m_2+m_1\},\nonumber \\
	&& b=\frac{1}{6}(3 m_3-2m_2-m_1).\nonumber \\
	{\rm{IO}} &:& p=\frac{1}{3}(2 m_1+m_2),\hspace{3mm} q=\frac{1}{6} \{(2 m_1+2 m_2)+(2\sqrt{2} m_1+\sqrt{2}m_2)s_{13}\},\nonumber \\
	&& r=\frac{1}{6} \{(2 m_1-2 m_2)+(2\sqrt{2} m_1+\sqrt{2}m_2)s_{13}\},\hspace{3mm}s=\frac{1}{6}(3 m_3+2 m_2+m_1),\nonumber \\
	&& b=\frac{1}{6}(3 m_3-2m_2-m_1).
	\end{eqnarray}
	Now putting all the values from neutrino oscillation data \cite{deSalas:2020pgw} in the expansion of the above equation for both NO and IO, all the numerical values of the Yukawa couplings $Y^{\nu}$ can be found, which is the topic of the following subsection.
	\subsection{Numerical analysis}\label{subsec:numerical analysis}
	Here we numerically evaluate the Yukawa coupling parameters $Y^{\nu}$ to satisfy the neutrino oscillation data. As we have fixed $\theta_{23}$, $\theta_{12}$ and $\delta_{\rm{CP}}$ in our analysis, in this regard, we vary the solar and atmospheric mass squared differences and $\sin^2\theta_{13}$ within the $3 \sigma$ region for both NO and IO, which are given below. \cite{deSalas:2020pgw}.
	\begin{eqnarray}
	&&	{\rm NO}:\hspace{4mm} m_s^2=(6.95- 8.14) \times 10^{-5}\hspace{2mm}{\rm eV^2},\hspace{3mm}m_a^2=(2.47-2.63)\times 10^{-3}\hspace{2mm}{\rm eV^2},\nonumber \\
	&&{\rm IO}:\hspace{4mm} m_s^2=(6.95- 8.14) \times 10^{-5}\hspace{2mm}{\rm eV^2},\hspace{3mm}m_a^2=(2.37-2.63)\times 10^{-3}\hspace{2mm}{\rm eV^2},\nonumber \\
	&& {\rm NO}:\hspace{3mm}\sin^2\theta_{13}=(0.02000-0.02405),\hspace{3mm}{\rm IO}:\hspace{3mm}\sin^2\theta_{13}=(0.02018-0.02424).
	\end{eqnarray}
	There is another type of NO of neutrino masses possible. The lightest mass $m_1$ is zero and the solar and atmospheric mass squared difference $m_s^2$ and $m_a^2$ are directly fitted into $m_2^2$ and $m_3^2$ in the diagonalization.
	
\begin{figure}[h!]
		\includegraphics[height=2in,width=3in]{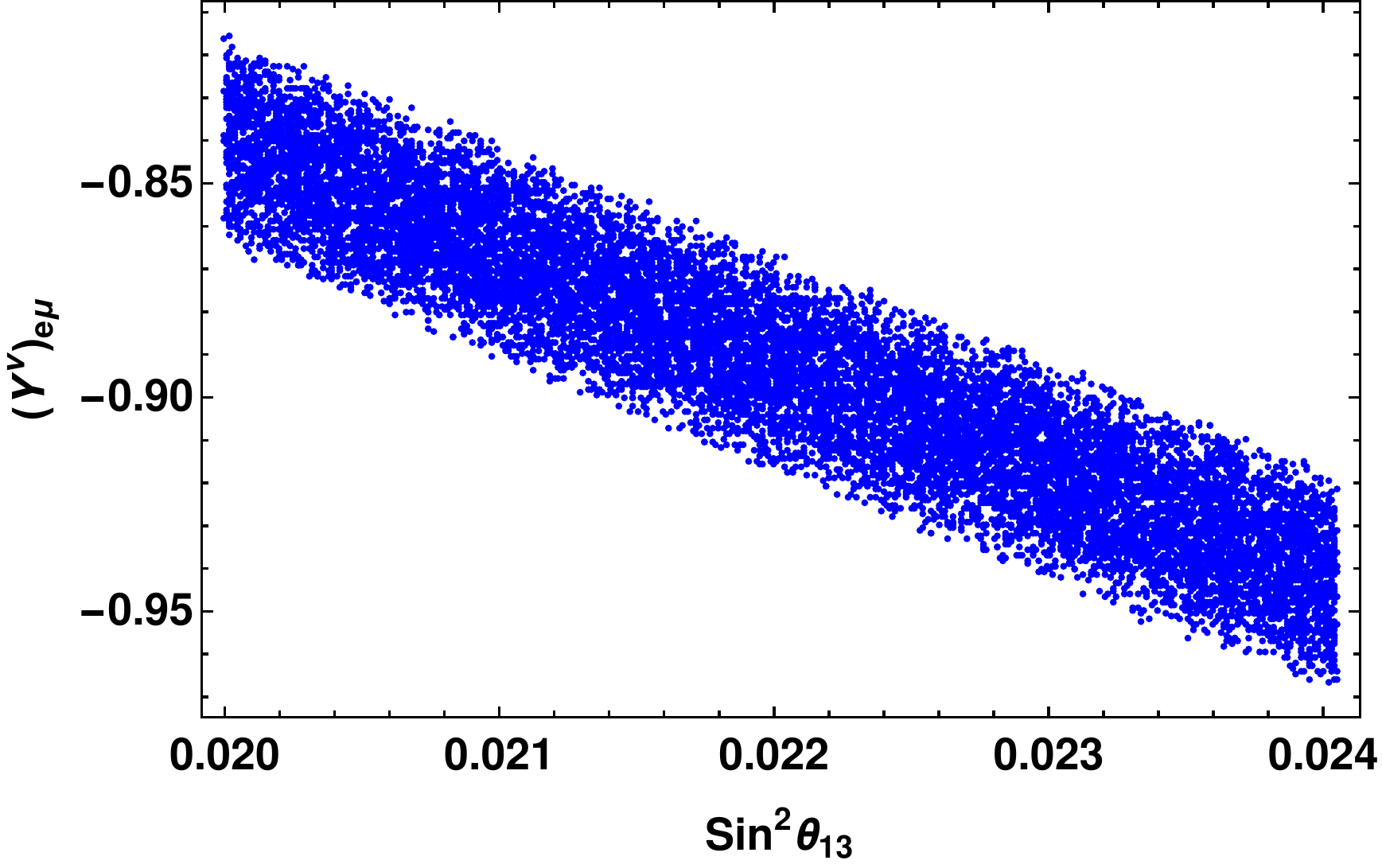}
		\includegraphics[height=2in,width=3in]{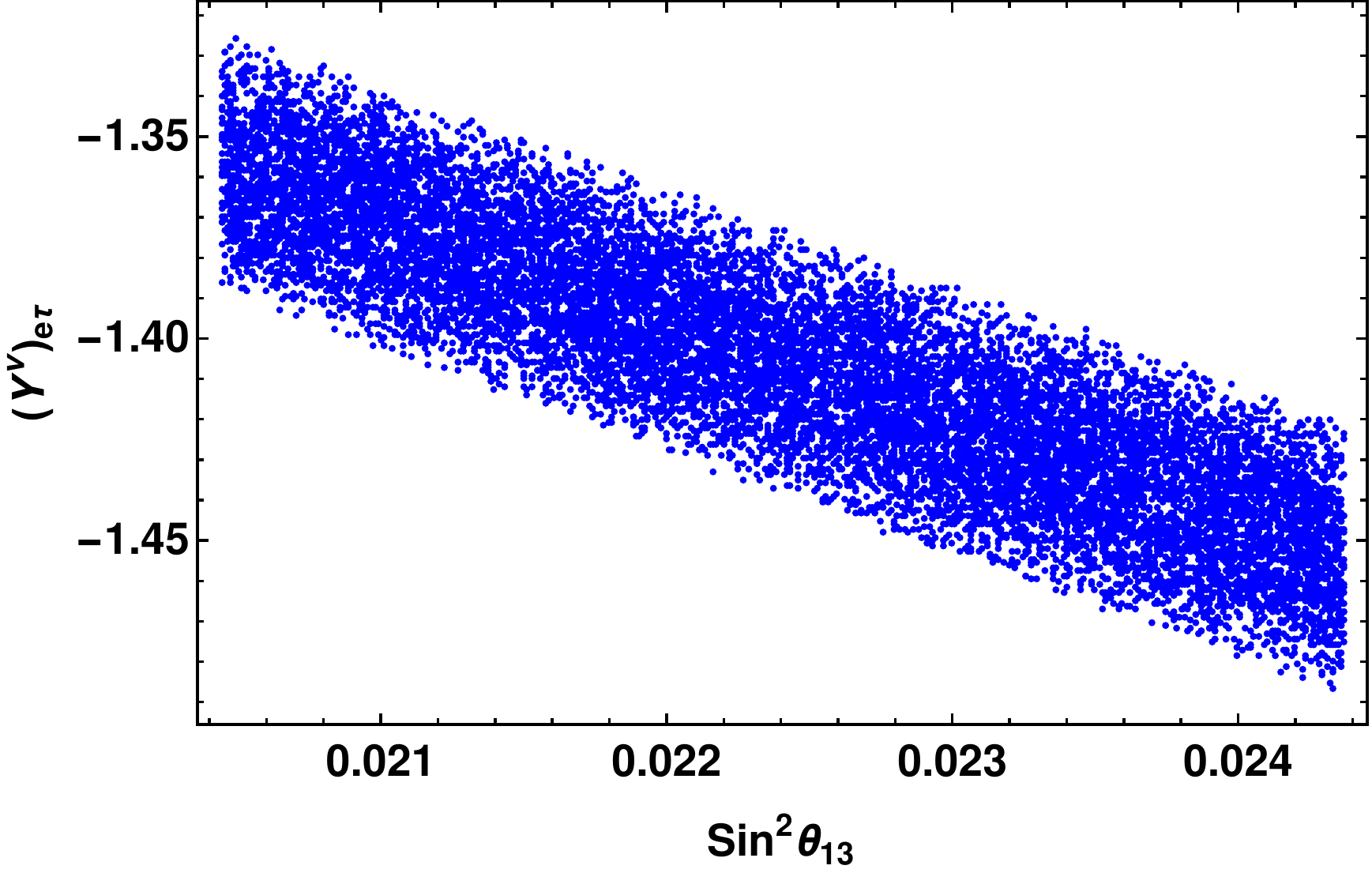}
		\caption{Variation of $Y_{e\mu}^{\nu}$(left) and $Y_{e\tau}^{\nu}$ (right) with the $3\sigma$ ranges of $\sin^2\theta_{13}$ for NO of neutrino masses. The ranges are coming out to be $\mathcal{O}(1)$ matching with our assumption.}
		\label{fig:NO}
	\end{figure}

\begin{figure}[h!]
		\includegraphics[height=2in,width=3in]{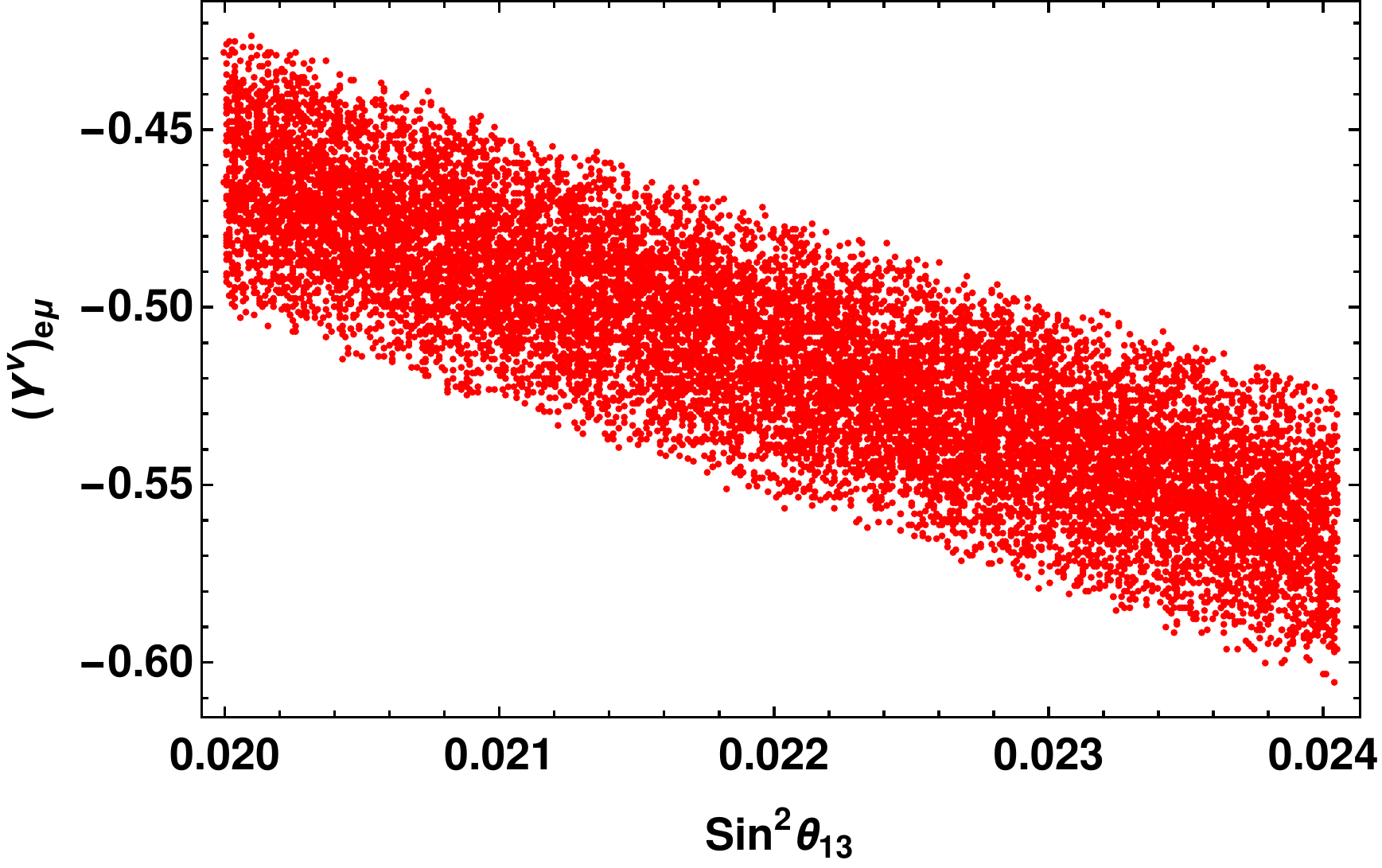}
		\includegraphics[height=2in,width=3in]{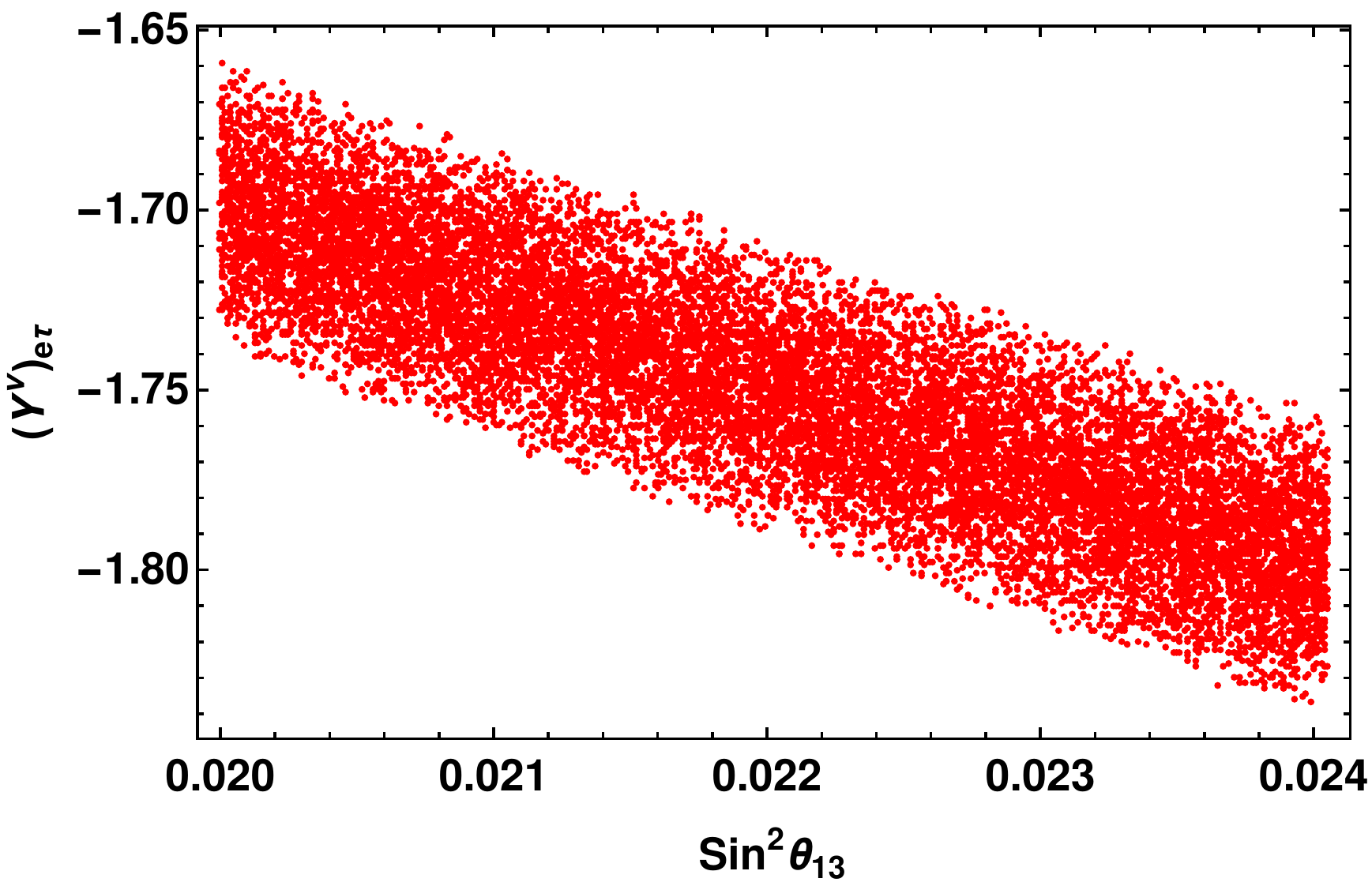}
		\caption{Variation of $Y_{e\mu}^{\nu}$(left) and $Y_{e\tau}^{\nu}$ (right) with the $3\sigma$ ranges of $\sin^2\theta_{13}$ for NO with $m_1=0$ of neutrino masses. The ranges are coming out to be $\mathcal{O}(1)$ matching with our assumption.}
		\label{fig:NO_m1-0}
	\end{figure}

\begin{figure}[h!]
		\includegraphics[height=2in,width=3in]{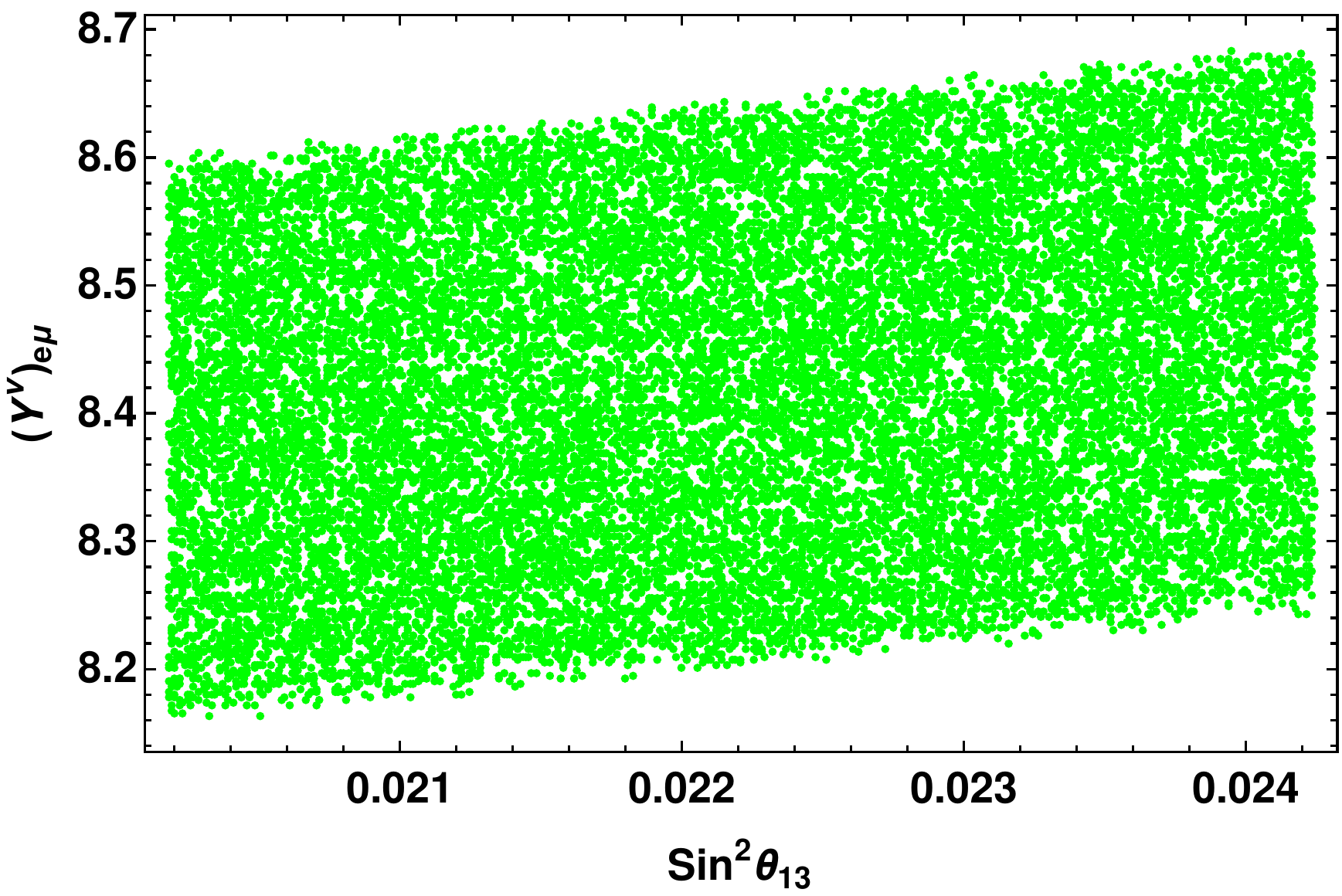}
		\includegraphics[height=2in,width=3in]{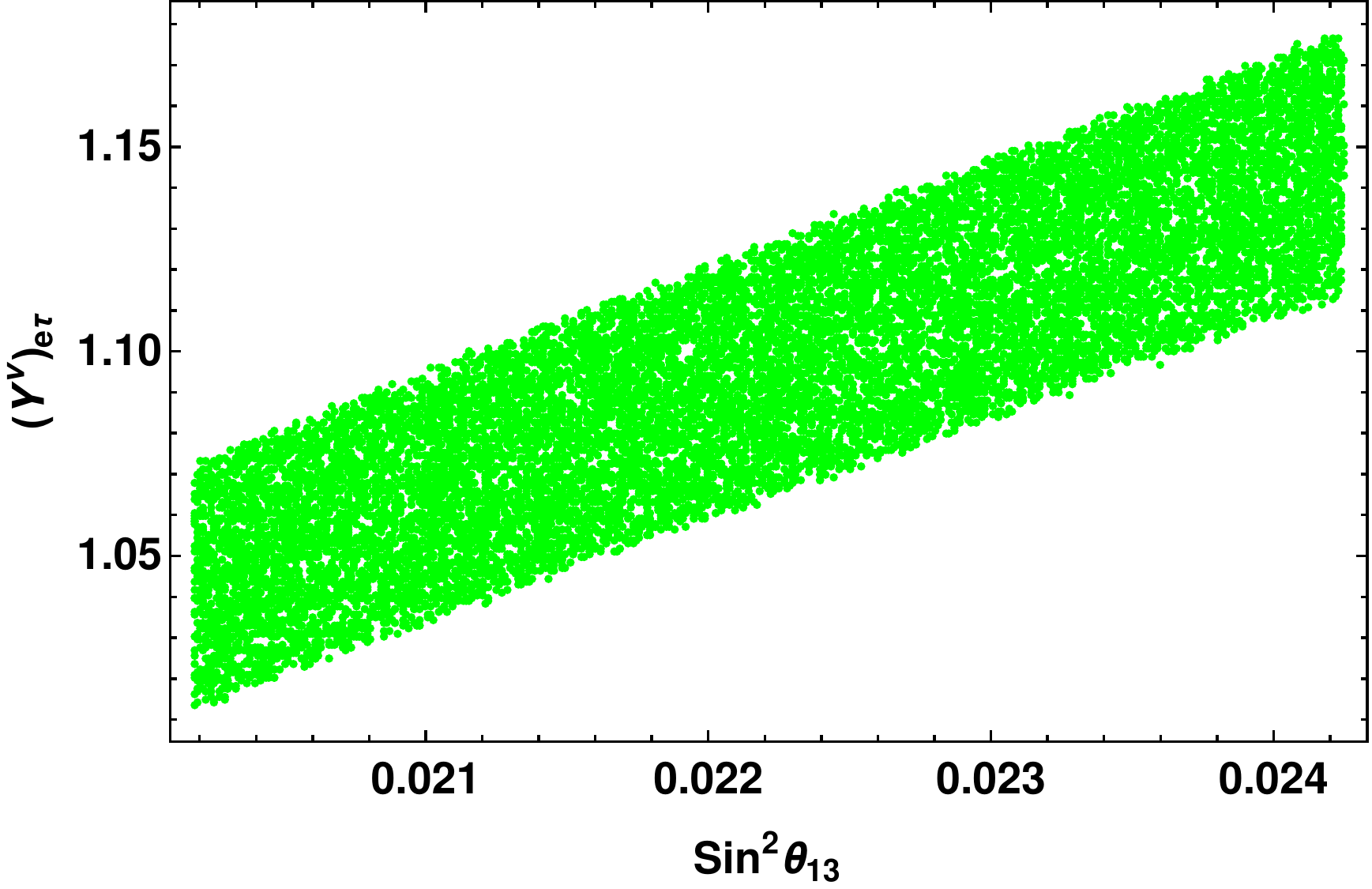}
		\caption{Variation of $Y_{e\mu}^{\nu}$(left) and $Y_{e\tau}^{\nu}$ (right) with the $3\sigma$ ranges of $\sin^2\theta_{13}$ for IO of neutrino masses. The ranges are not coming out to be $\mathcal{O}(1)$ discarding our assumption.}
		\label{fig:IO}
	\end{figure}

	In our analysis, we take $\delta=\epsilon=1/6$ to get the results. In doing the calculation to satisfy the neutrino oscillation data for neutrino masses we take $v_{\Delta}=1$ eV. This small VEV arises in this model for some fine-tuning of the  parameters in the scalar potential given in section \ref{sec:scalar potential}. As it is seen from Eq. (\ref{eq:expansion_mnu/ma}) that up to first order in $\sin \theta_{13}$, $Y_{ee}^{\nu}$, $Y_{\mu\mu}^{\nu}$, $Y_{\mu \tau}^{\nu}$ and $Y_{\tau \tau}^{\nu}$ do not depend on $\sin \theta_{13}$. One thing to notice here is that the approximate expressions for $Y_{\mu\mu}^{\nu}$ and $Y_{\tau \tau}^{\nu}$ are same can be seen from Eq. (\ref{eq:expansion_mnu/ma}). Now with the above numerical values for mass squared differences, ranges for these $Y^{\nu}$ can be found, which are summarized in the Table. [\ref{tab:table1}].
	\begin{table}[h!]
		\begin{center}

			\begin{tabular}{|c|c|c|c|c|c|c|}
				\hline
				\multirow{3}{*}{} & \multicolumn{2}{c|}{NO} & %
				\multicolumn{2}{c|}{IO} &  \multicolumn{2}{c|}{NO($m_1=0$)}\\
				\cline{2-7}
				& Min & Max & Min & Max & Min & Max\\
				\hline
				$Y_{ee}$ & 2.048 & 2.418 & 10.72 & 11.30 & 0.599 & 0.649 \\
				\hline
				$Y_{\mu \mu}$ & 1.098 & 1.144 & 1.047 & 1.10 & 0.994 & 1.03 \\
				\hline
				$Y_{\mu \tau}$ & 0.702 & 0.744 & -0.793 & -0.738 & 0.786 & 0.823 \\
				\hline
				$Y_{\tau\tau}$ & 1.098 & 1.144 & 1.047 & 1.10 & 0.994 & 1.03 \\
				\hline	
			\end{tabular}
		\end{center}
		\caption{Allowed ranged for the Yukawa couplings by fitting the neutrino oscillation data. Here $Y_{\mu\mu}$ and $Y_{\tau\tau}$ are same.}
		\label{tab:table1}
	\end{table}
	Looking at the Eq. (\ref{eq:expansion_mnu/ma}), the Yukawa couplings $Y_{e\mu}$ and $Y_{e\tau}$ depend on masses as well as $\theta_{13}$. Now taking care of this $3\sigma$ ranges of solar and atmospheric mass squared differences, the variation of these couplings with respect to the $3\sigma$ ranges of $\sin^2\theta_{13}$ are shown in figures \ref{fig:NO}, \ref{fig:NO_m1-0}, \ref{fig:IO} for NO, NO with $m_1=0$ and IO respectively.
	Now let us summarize the particular values of all the couplings by fitting these to the best-fit values of the neutrino oscillation data. For NO and IO, the couplings which fits the neutrino oscillation data are
	\begin{eqnarray}\label{NO LO Yukawa values}
	&&\rm{NO}:\hspace{4mm}	(Y_{ee}^{\nu}, Y_{e\mu}^{\nu}, Y_{e\tau}^{\nu}, Y_{\mu \mu}^{\nu}, Y_{\mu \tau}^{\nu}, Y_{\tau \tau}^{\nu}) \approx (2.12,-0.91, -1.43,1.12, 0.72,1.12). \nonumber \\
	&& \rm{IO}: \hspace{4mm}(Y_{ee}^{\nu}, Y_{e\mu}^{\nu}, Y_{e\tau}^{\nu}, Y_{\mu \mu}^{\nu}, Y_{\mu \tau}^{\nu}, Y_{\tau \tau}^{\nu}) \approx (10.90,8.35,1.09, 1.06, -0.757,1.06). 
	\end{eqnarray}
	Above results indicate that for the IO of neutrino masses, the Yukawa couplings are coming out to be large compared to unity but the $\mathcal{O}(1)$ approximation for these couplings are well valid for NO. We can check our result for the normal ordering with $m_1=0$, $m_2=m_s=0.0087$ eV and $m_3=m_a=0.05$ eV, where these couplings now are found to be
	\begin{eqnarray}\label{NO m1=0 Yukawa values}
	(Y_{ee}^{\nu}, Y_{e\mu}^{\nu}, Y_{e\tau}^{\nu}, Y_{\mu \mu}^{\nu}, Y_{\mu \tau}^{\nu}, Y_{\tau \tau}^{\nu}) \approx (0.62,-0.53, -1.78,1.01, 0.80, 1.01).
	\end{eqnarray}
	Here the $\mathcal{O}(1)$ approximation is valid for NO of neutrino masses where the lightest neutrino mass is zero. Hence, looking at Eqs. (\ref{NO LO Yukawa values}), (\ref{NO m1=0 Yukawa values}), we conclude that our model predicts NO of neutrino masses and discards IO of neutrino masses.
	
	
	In principle, one could generate the type II seesaw Lagrangian of Eq. (\ref {neutrino lagrangian}) by $X$ field only, but there are two problems: If one uses $X^{\dagger}X$ in place of ${X^{\prime}}^2$ then the theory cannot be UV complete. The entire UV complete Lagrangian of Eq. (\ref{neutrino lagrangian}) is described in detail in the next subsection. The second one is that only one singlet field can be used in the Lagrangian of Eq. (\ref {neutrino lagrangian}) if we omit the $Z_3$ symmetry from the model, then also it is possible to explain neutrino mass and mixing with real Yukawa couplings, but then the problem is with UV completion. We will not be able to generate $\frac{X^3}{M^3}$ or $\frac{X^2}{M^2}$ simultaneously even if we use different vector-like leptons while writing UV complete Lagrangian described elaborately in the following subsection.
	
	\subsection{UV completion}\label{subsec:UV completion}
	Here we present the UV completion mechanism for the model of neutrino mixing to illustrate the origin of higher-dimensional terms in Eq. (\ref {neutrino lagrangian}). Our method of UV completion is similar to the Refs.  \cite{Lykken:2008bw,Ghosh:2011up,Grossmann:2010ea} but those mechanisms are used for type-I seesaw, while there is no UV completion mechanism for type II seesaw, which we have tried to present here. We add two symmetries $U(1)_X \times U(1)_F$ apart from some heavy vector-like fermions and flavon fields $F_i$. After that, we can write the renormalizable Lagrangian between SM fields and extra fields, and upon integrating the heavy fields, we can get the higher dimensional Lagrangian of Eq. (\ref {neutrino lagrangian}). The singlet field $X$ has zero $U(1)_F$ charge, and flavon fields are singlets under $U(1)_X$. $U(1)_X$ symmetry is broken when $X$ and $X^{\prime}$ fields acquire VEV at some lower scale. The VEV of flavon fields $F_i$ will break the $U(1)_F$ field spontaneously. In the Lagrangian of Eq. (\ref{neutrino lagrangian}), two types of higher dimensional terms are involved which are $\Big(\frac{X}{M}\Big)^3$ and $\Big(\frac{X^{\prime}}{M}\Big)^2$. Hence, 
	the motivation for using two singlets in the neutrino sector is to consistently generate the UV complete Lagrangian. Our model is composed of $CP \times Z_3$, so it is needed to make the extra fields invariant under this symmetry.

	To generate the term of having coefficient $Y_{ee}^{\nu}$ of Eq. (\ref{neutrino lagrangian}), we add three pairs of vector-like fermions $(D_{iL}, D_{iR})$, $i$=1, 2, 3, which have the same hypercharge as the lepton doublets of the standard model apart from one flavon field $F_1$. 
	The CP symmetry can be written in a similar way as Eq. (\ref{CP transformation}).
	\begin{eqnarray}\label{UV completion CP assign}
	CP: \hspace{3mm} D_{i L} \rightarrow i \gamma^0 C \bar{D}_{i L}^T, \hspace{3mm}D_{i R} \rightarrow i \gamma^0 C \bar{D}_{i R}^T,\hspace{3mm X\rightarrow} X^*, \hspace{3mm} F_1 \rightarrow F_1^*, \hspace{3mm}\Delta \rightarrow	\Delta^*.
	\end{eqnarray}
	Here $i=1,2,3$.
	
	The invariant Lagrangian is written in the following way
	\begin{eqnarray}\label{Lag UV completion 1}
	&&\mathcal{L}_{ee}^{\nu}= \bar{D}_{eL}^c i \sigma_2 \Delta D_{1L} + \bar{D}_{1L}D_{1R}F_1+ \bar{D}_{1R}D_{2L}X+  \bar{D}_{2L}D_{2R}F_1 +  \bar{D}_{2R}D_{3L}X + \bar{D}_{3L}D_{3R}F_1 \nonumber \\
	&&+ \bar{D}_{3R}D_{eL}X.
	\end{eqnarray}
	The above lagrangian is invariant under $CP$ symmetry. So, the couplings associated with each term will be real which we have not explicitly written in the above equation. These couplings are of ${\mathcal{O}}(1)$ parameter.
	%
	Now when the flavon field $F_1$ acquires VEV and then the $U(1)_F$ symmetry is broken spontaneously around the scale $M$. The VEV of these flavons gives rise to masses of the flavon fields around the scale $M$. Then, the term of $Y_{ee}^{\nu}$ of Eq. (\ref{neutrino lagrangian}) will be originated after integrating out the heavy vector-like fermions and flavons of Eq. (\ref{Lag UV completion 1}). 
	
	The other terms of Eq. (\ref{neutrino lagrangian}) can be generated analogously by introducing vector-like leptons $D_{iL},D_{iR}$, where $i=4\cdots 14$, and flavon fields $F_2$, $F_3$ and $F_4$. The properties of these vector-like fermions and flavon fields are the same as in the previous paragraph. The CP symmetry on these fields is exactly the same as in Eq. (\ref {UV completion CP assign}). $CP$ symmetry on $X^{\prime}$ is also the same as $X$ transformation. Hence, the invariant renormalizable Lagrangian, which generates all the terms in the neutrino Lagrangian of Eq. (\ref{neutrino lagrangian}) can be written in the following way, where the charge assignments of vector-like leptons, flavon fields, $X$ and $X^{\prime}$ under $U(1)_X \times U(1)_F \times Z_3$ are shown in Eqs. (\ref{eq:U(1)X charge assign}), (\ref{eq:U(1)F charge assign}) and (\ref{eq:Z3 charge assign}).
	\begin{eqnarray}\label{UV completion Lag 2}
	&&{\mathcal{L}}_{e\mu}^{\nu}=\bar{D}_{eL}^c i\sigma_2\Delta D_{1L}+\bar{D}_{1L}D_{1R}F_1+\bar{D}_{1R}D_{4L}X+\bar{D}_{4L}D_{4R}F_1^*+\bar{D}_{4R}D_{5L}X+\bar{D}_{5L}D_{5R}F_1^* \nonumber \\
	&&+\bar{D}_{5R}D_{\mu L}X+h.c.,
	\end{eqnarray}
	\begin{eqnarray}
	&&{\mathcal{L}}_{e\tau}^{\nu}=\bar{D}_{eL}^c i\sigma_2\Delta D_{6L}+\bar{D}_{6L}D_{6R}F_1^*+\bar{D}_{6R}D_{7L}X+\bar{D}_{7L}D_{7R}F_1^*+\bar{D}_{7R}D_{8L}X+\bar{D}_{8L}D_{8R}F_1^* \nonumber \\
	&&+\bar{D}_{8R}D_{\tau L}X+h.c.,
	\end{eqnarray}
	\begin{eqnarray}
	{\mathcal{L}}_{\mu\mu}^{\nu}=\bar{D}_{\mu L}^c i\sigma_2\Delta D_{9L}+\bar{D}_{9L}D_{9R}F_2+\bar{D}_{9R}D_{10L}X^{\prime}+\bar{D}_{10L}D_{10R}F_2+\bar{D}_{10R}D_{\mu L}X^{\prime}+h.c.,
	\end{eqnarray}
	\begin{eqnarray}
	{\mathcal{L}}_{\tau\tau}^{\nu}=\bar{D}_{\tau L}^c i\sigma_2\Delta D_{11L}+\bar{D}_{11L}D_{11R}F_3+\bar{D}_{11R}D_{12L}X^{\prime}+\bar{D}_{12L}D_{12R}F_3+\bar{D}_{12R}D_{\tau L}X^{\prime}+h.c.,
	\end{eqnarray}
	\begin{eqnarray}
	{\mathcal{L}}_{\mu\tau}^{\nu}=\bar{D}_{\mu L}^c i\sigma_2\Delta D_{13L}+\bar{D}_{13L}D_{13R}F_4+\bar{D}_{13R}D_{14L}X^{\prime}+\bar{D}_{14L}D_{14R}F_4+\bar{D}_{14R}D_{\tau L}X^{\prime}+h.c..
	\end{eqnarray}
	All UV complete Lagrangian can be made invariant under $U(1)_X$ by the following charge assignment of all the fields by taking the charge of $X$ field $x$ under $U(1)_X$ to be independent.
	\begin{eqnarray}\label{eq:U(1)X charge assign}
	U(1)_X&:& D_{eL},D_{\mu L}, D_{\tau L}\rightarrow -\frac{3}{2}x,\quad F_1,F_2,F_3,F_4 \rightarrow 0,\nonumber \\ && D_{1L},D_{1R},D_{6L},D_{6R},D_{9L},D_{9R},D_{11L},D_{11R},D_{13L},D_{13R},X^{\prime}\rightarrow \frac{3}{2}x,,\nonumber \\
	&& D_{2L},D_{2R},D_{4L},D_{4R},D_{7L},D_{7R}\rightarrow \frac{1}{2}x,D_{3L},D_{3R},D_{5L},D_{5R},D_{8L},D_{8R}\rightarrow -\frac{1}{2}x,\nonumber \\  &&D_{10L},D_{10R},D_{12L},D_{12R},D_{14L},D_{14R}\rightarrow 0.
	\end{eqnarray}
	Again by taking charge of $F_1$ under $U(1)_F$, $f$ to be independent, charge assignment of all other fields are shown below.
	\begin{eqnarray}\label{eq:U(1)F charge assign}
	U(1)_F&:&
	D_{eL},D_{3R}\rightarrow -\frac{3}{2}f,\quad D_{\mu L},D_{5R},D_{6R},D_{7L}\rightarrow \frac{5}{2}f,\quad \nonumber \\
	&&D_{\tau L},D_{8R},D_{12R},D_{14R}\rightarrow \frac{9}{2}f,\quad D_{1L},D_{4R},D_{5L},D_{6L}\rightarrow \frac{3}{2}f\nonumber \\
	&& D_{1R},D_{2L},D_{4L}\rightarrow\frac{1}{2}f,\quad D_{2R},D_{3L}\rightarrow -\frac{1}{2}f,\quad D_{7R},D_{8L}\rightarrow \frac{7}{2}f\nonumber \\
	&& F_2,D_{9L},D_{10R},D_{13L}\rightarrow -\frac{5}{2}f,\quad D_{9R}, D_{10L},D_{11R},D_{12L}\rightarrow 0,\nonumber \\
	&& D_{13R},D_{14L}\rightarrow f,\quad F_3\rightarrow -\frac{9}{2}f,\quad F_4 \rightarrow -\frac{7}{2}f,\quad X,X^{\prime}\rightarrow 0.
	\end{eqnarray}
	The $Z_3$ charge assignment of all the fields are
	\begin{eqnarray}\label{eq:Z3 charge assign}
	Z_3&:&
	X,D_{3R},D_{3L},D_{5L},D_{5R},D_{8L},D_{8R},\rightarrow \omega,\quad D_{2R},D_{2L},D_{4L},D_{4R},D_{7L},D_{7R}\rightarrow \omega^2, \nonumber \\
	&& X^{\prime},F_1,F_2,F_3,F_4,D_{eL},D_{\mu L},D_{\tau L},D_{1L},D_{1R},D_{6L},D_{6R},D_{9L},D_{9R},D_{10L},\nonumber \\
	&& D_{10R},D_{11L},D_{11R},D_{12L},D_{12R},D_{13L},D_{13R},D_{14L},D_{14R} \rightarrow 1.
	\end{eqnarray}
	With this UV complete mechanism, we can get the terms given in Eq. (\ref {neutrino lagrangian}). There are terms like $\frac{X^{\prime}}{M^2}\bar{D}_{eL}^ci\sigma_2\Delta D_{eL} $ which is allowed by $U(1)_X$ symmetry but disallowed by the UV completion mechanism, which means this term cannot be generated from a renormalized Lagrangian with the given vector-like fermions and flavon fields.
	
	As we have seen from the charge assignment of Eqs. (\ref{eq:U(1)X charge assign}), (\ref{eq:U(1)F charge assign}) and (\ref{eq:Z3 charge assign}), that $X$ and $X^{\prime}$ have different $U(1)_X$ charge. This is the reason why the same singlet field is not used in the neutrino Lagrangian of Eq. (\ref {neutrino lagrangian}). It will not be possible to write the UV complete Lagrangian simultaneously for $Y_{ee}^{\nu}$ and $Y_{e\mu}^{\nu}$ with only one singlet field.
	
	We want to mention one important thing that in this work, the Yukawa couplings are real, and the CP-violating phase is $\pi$ which is allowed by the recent neutrino oscillation data \cite{deSalas:2020pgw}. In principle, we can explain the CP-violating phase other than $\pi$ in the same model if the Yukawa couplings are complex. In that case, the complex VEV of the singlet field $X$ can be the source of the complex Yukawa couplings, and then the UV completion process described above will be different. This possibility of explaining $CP$ violation is explained in Section \ref{sec:cp violation model}.
	\section{Lepton flavor violation}\label{sec:LFV}
	In this model, the neutrino mixing has been generated via type II seesaw, which involves triplet scalar $\Delta$. So one singly charged Higgs ($\Delta^+$) and one doubly charged Higgs ($\Delta^{++}$) will come from this triplet. As a model prediction we calculate the branching fraction of lepton flavor violating decays $\mu \rightarrow e \gamma$ and $\mu \rightarrow \bar{e} e e$ \cite{Lindner:2016bgg,Akeroyd:2009nu,Garcia-Aguilar:2021xgk}.

	The branching ratio for $\mu \rightarrow e\gamma$ is given by \cite{Akeroyd:2009nu}
	\begin{eqnarray}\label{BR mu}
	{\rm Br}(\mu \rightarrow e\gamma) \approx 4.5 \times 10^{-3}\Bigg(\frac{1}{ v_{\Delta}}\Bigg)^4\Big|(V_L^{\nu}D_m^{\dagger}D_m {V_L^{\nu}}^{\dagger})_{e\mu}\Big|^2 \Bigg(\frac{200 \hspace{1mm}{\rm GeV}}{m_{\Delta}}\Bigg)^4.
	\end{eqnarray}
	Here we are assuming that $m_{\Delta^{++}}=m_{\Delta^+}=m_{\Delta}$. $D_m={\rm diag}(m_1, m_2, m_3)$ is the diagonal neutrino mass matrix, and $V_{L}^{\nu}$ is defined in Eq. (\ref{V l nu}). In the previous section, all the couplings were calculated using the best-fit values of neutrino oscillation data. As a result we use $s_{13} \approx 0.15$ in $V_{L}^{\nu}$. Again from the previous section, $\epsilon$ and $v_{\Delta}$ are fixed to $1/6$ and 1 eV, respectively. The mass eigenvalues in $D_m$ can be two types, NO and IO, but we will calculate the branching fraction for NO as our calculation prefers NO of neutrino masses. The experimental upper limit is ${\rm Br}(\mu \rightarrow e\gamma)< 4.2 \times 10^{-13}$ \cite{MEG:2016leq}. It gives a lower limit on the mass of doubly and singly charged triplet in our model, which is $m_{\Delta}>1$ TeV.
	
	Our model predicts the branching fraction for the lepton flavor violating decay $\mu \rightarrow 3 e$, which can probe the physics beyond the standard model. In the Higgs triplet model, this decay has been mediated by $\Delta^{\pm \pm}$. Within the experimental limit of the LFV decays, the mass of the doubly charged Higgs can be constrained in this model after using the couplings of Eq. (\ref{NO LO Yukawa values}). The branching ratio for $\mu \rightarrow 3 e$ \cite{Akeroyd:2009nu} is given by
	\begin{eqnarray}
	{\rm Br}(\mu \rightarrow 3 e)\approx 1.1 \Bigg(\frac{1}{\epsilon}\Bigg)^{12}|Y_{e\mu}^{\nu}|^2 |Y_{ee}^{\nu}|^2 \Bigg(\frac{200 \hspace{1mm}{\rm GeV}}{m_{\Delta}^{++}}\Bigg)^4.
	\end{eqnarray}
	As the values of $\epsilon$, $Y_{e\mu}^{\nu}$ and $Y_{ee}^{\nu}$ are already fixed by the neutrino oscillation data in the previous section, then similar to the $\mu \rightarrow e\gamma$ case, we can find a lower limit on the mass of doubly charged Higgs. Putting the experimental upper limit on ${\rm Br}(\mu \rightarrow 3 e)< 1.0 \times 10^{-12}$ \cite{Bell}, we calculated the lower limit on the mass of $\Delta^{\pm \pm}$ as $m_{\Delta^{\pm \pm}}>1.2$ TeV. 
	\begin{figure}[h!]\label{fig:LFV NH}
		\includegraphics[height=2.0in,width=3.0in]{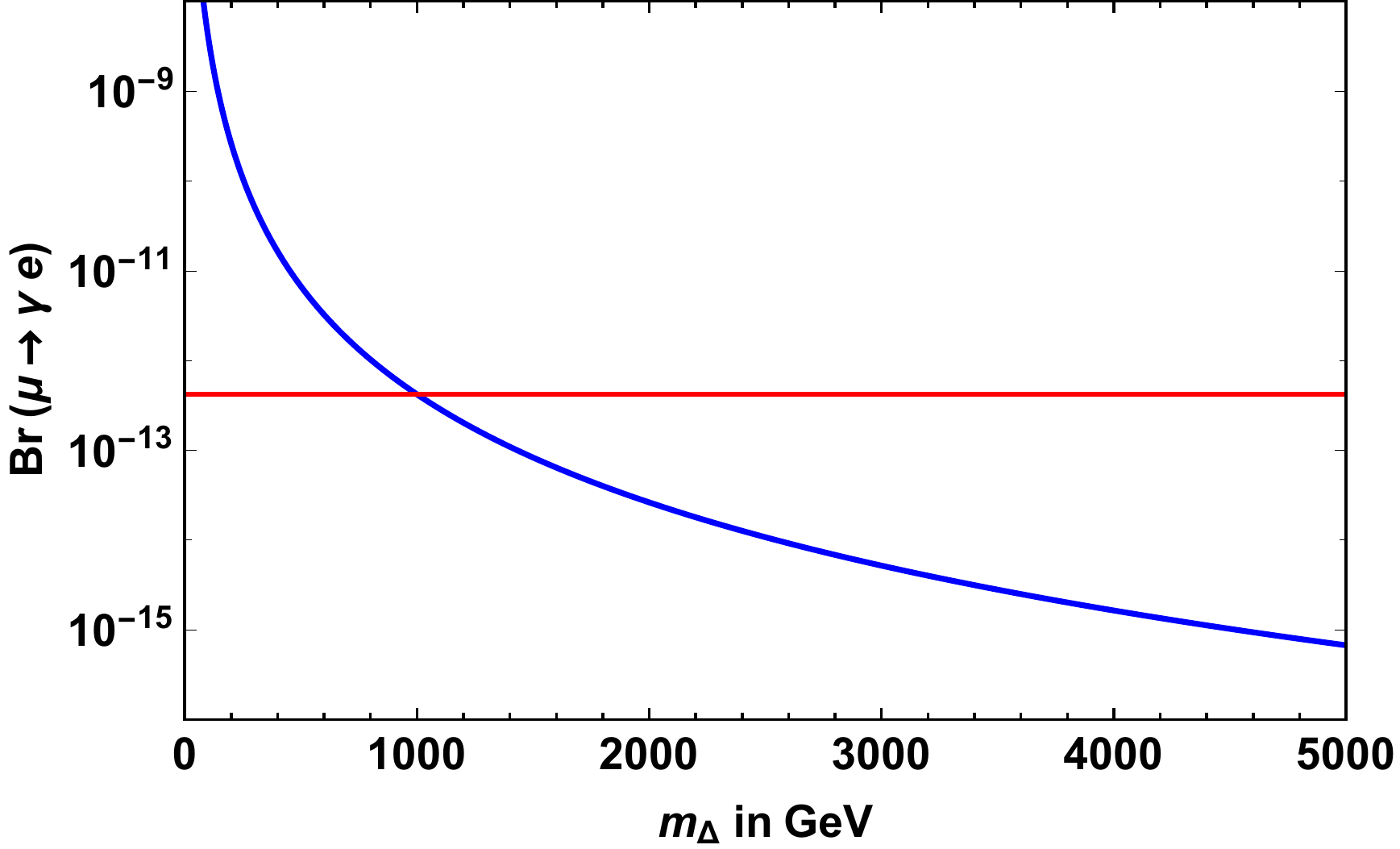}
		\includegraphics[height=2.0in,width=3.0in]{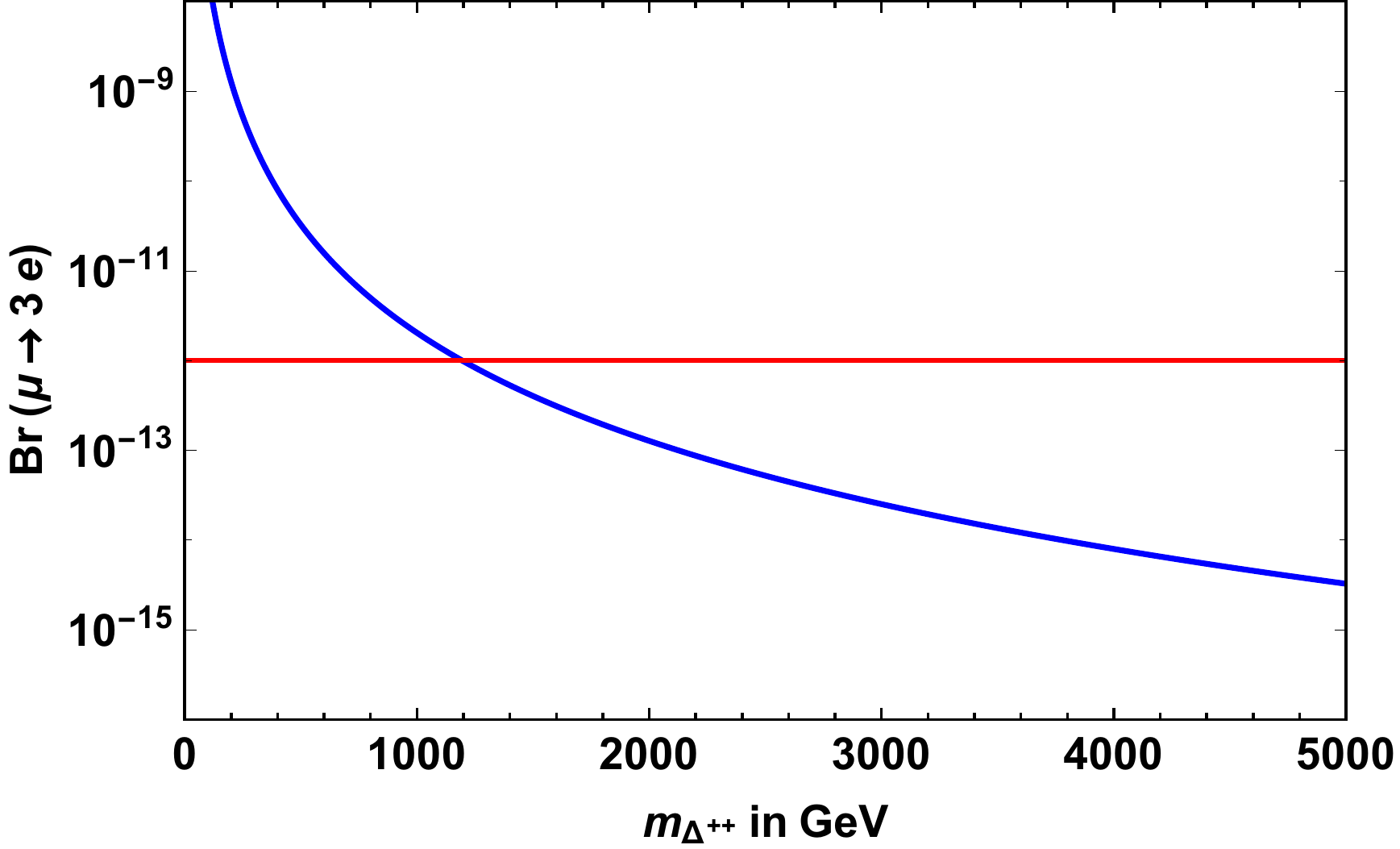}
		\caption{From left to right: ${\rm Br}(\mu\rightarrow e\gamma)$ versus $m_{\Delta}$ parameter and ${\rm Br}(\mu\rightarrow 3 e)$ versus $m_{\Delta^{\pm \pm}}$ where masses have been taken in TeV unit. Red line corresponds to the experimental upper limit while the blue line shows the prediction of our model. The region above red line is excluded while below the red line is allowed for the masses of the triplet scalar.}
		\label{fig:branching}
	\end{figure}
	In  the Fig. \ref{fig:branching}, branching fraction for both $\mu \rightarrow e\gamma$ and $\mu \rightarrow 3 e$ are plotted as a function of mass of the triplet $m_{\Delta}$ and mass of doubly charged Higgs ($m_{\Delta^{\pm \pm}}$) as all the other parameters are fixed in the previous section. This model predicts that to satisfy the experimental upper bound on the branching fraction for both the decays, the mass of the singly charged Higgs ($\Delta^{\pm}$) should be greater than 1 TeV and doubly charge Higgs ($\Delta^{\pm \pm}$) must have mass greater than 1.2 TeV.  
	\section{ Scalar potential}\label{sec:scalar potential}
	In this section, we analyze the scalar potential of our model, which involves many scalars to generate fermion masses described in the previous sections. The full invariant potential of this model is given below.
	\begin{eqnarray}
	V& =&-m_{\phi_1}^2 \phi_1^{\dagger}\phi_1- m_{\phi_2}^2 \phi_2^{\dagger}\phi_2  -m_{X}^2 (X^{\dagger}X)-m_{X^{\prime}}^2 ({X^{\prime}}^{\dagger}X^{\prime})- m_{F_1}^2 (F_1^{\dagger}F_1) -m_{F_2}^2(F_2^{\dagger}F_2)\nonumber \\
	&& -m_{F_3}^2 (F_3^{\dagger}F_3)-m_{F_4}^2 (F_4^{\dagger}F_4) - m_{\Delta}^2 {\rm tr}(\Delta^{\dagger}\Delta)+\lambda_{1} (\phi_1^{\dagger}\phi_1)^2+ \lambda_{2} (\phi_2^{\dagger}\phi_2)^2 + \lambda_{\Delta}\Big({\rm  tr}(\Delta^{\dagger}\Delta)\Big)^2 \nonumber \\
	&& +\lambda_{\Delta}^{\prime}{\rm Tr}(\Delta^{\dagger}
	\Delta^{\dagger}){\rm Tr}(\Delta\Delta) +\lambda_{X}(X^{\dagger}X)^2 +\lambda_{X^{\prime}}({X^{\prime}}^{\dagger}X^{\prime})^2  +\lambda_{F_1}(F_1^{\dagger}F_1)^2+ \lambda_{F_2}(F_2^{\dagger}F_2)^2 \nonumber \\
	&&+\lambda_{F_3}(F_3^{\dagger}F_3)^2+\lambda_{F_4}(F_4^{\dagger}F_4)^2+ \lambda_{11}[(\phi_1^{\dagger} \phi_2)^2+(\phi_2^{\dagger} \phi_1)^2] +\lambda_{12}  (\phi_1^{\dagger}\phi_2)(\phi_2^{\dagger}\phi_1) \nonumber \\
	&&+ i \lambda_{13} (\phi_1^{\dagger}\phi_1)(\phi_1^{\dagger}\phi_2-\phi_2^{\dagger}\phi_1)+ i \lambda_{14} (\phi_2^{\dagger}\phi_2)(\phi_1^{\dagger}\phi_2-\phi_2^{\dagger}\phi_1) +\lambda_{15} (\phi_1^{\dagger}\phi_1)(\phi_2^{\dagger}\phi_2) \nonumber \\
	&& +\lambda_{16}(\phi_1^{\dagger}\phi_1){\rm{tr}} (\Delta^{\dagger}\Delta)+ \lambda_{17}(\phi_2^{\dagger}\phi_2){\rm tr}(\Delta^{\dagger}\Delta) +  i \lambda_{18}(\phi_1^{\dagger}\phi_2 -\phi_2^{\dagger}\phi_1){\rm tr}(\Delta^{\dagger}\Delta)  \nonumber \\
	&& +\lambda_{19}\phi_1^{\dagger}\Delta_1^{\dagger}\Delta_1 \phi_1 + \lambda_{20}\phi_2^{\dagger}\Delta^{\dagger}\Delta \phi_2  + +\lambda_{21}(X^{\dagger}X)(\phi_{1}^{\dagger}\phi_{1}) +\lambda_{22}(X^{\dagger}X)(\phi_{2}^{\dagger}\phi_{2}) \nonumber \\
	&& +i \lambda_{23}(X^{\dagger}X)(\phi_1^{\dagger}\phi_2-\phi_2^{\dagger}\phi_1)+ \lambda_{24}(X^{\dagger}X){\rm tr}(\Delta^{\dagger}\Delta)+\lambda_{25}({X^{\prime}}^{\dagger}X^{\prime})(\phi_{1}^{\dagger}\phi_{1}) \nonumber \\
	&&
	+\lambda_{26}({X^{\prime}}^{\dagger}X^{\prime})(\phi_{2}^{\dagger}\phi_{2})+i\lambda_{27}({X^{\prime}}^{\dagger}X^{\prime})(\phi_{1}^{\dagger}\phi_{2}-\phi_{2}^{\dagger}\phi_{1}) +\lambda_{28}({X^{\prime}}^{\dagger}X^{\prime}){\rm tr}(\Delta^{\dagger}\Delta) \nonumber \\
	&& \lambda_{29}(F_1^{\dagger}F_1)(\phi_{1}^{\dagger}\phi_{1})+\lambda_{30}(F_1^{\dagger}F_1)(\phi_{2}^{\dagger}\phi_{2})+\lambda_{31}(F_1^{\dagger}F_1)(\phi_{1}^{\dagger}\phi_{2}-\phi_{2}^{\dagger}\phi_{1})+\lambda_{32}(F_1^{\dagger}F_1){\rm Tr}(\Delta^{\dagger}\Delta)\nonumber \\
	&& 
	\lambda_{33}(F_1^{\dagger}F_1)(X^{\dagger}X)+\lambda_{34}(F_1^{\dagger}F_1)({X^{\prime}}^{\dagger}X^{\prime})+\lambda_{35}(F_2^{\dagger}F_2)(\phi_{1}^{\dagger}\phi_{1})+\lambda_{36}(F_2^{\dagger}F_2)(\phi_{2}^{\dagger}\phi_{2})\nonumber \\
	&&+\lambda_{37}(F_2^{\dagger}F_2)(\phi_{1}^{\dagger}\phi_{2}-\phi_{2}^{\dagger}\phi_{1})+\lambda_{38}(F_2^{\dagger}F_2){\rm Tr}(\Delta^{\dagger}\Delta)+
	\lambda_{39}(F_2^{\dagger}F_2)(X^{\dagger}X)\nonumber \\
	&& +\lambda_{40}(F_2^{\dagger}F_2)({X^{\prime}}^{\dagger}X^{\prime})+\lambda_{41}(F_3^{\dagger}F_3)(\phi_{1}^{\dagger}\phi_{1})+\lambda_{42}(F_3^{\dagger}F_3)(\phi_{2}^{\dagger}\phi_{2})+\lambda_{43}(F_3^{\dagger}F_3)(\phi_{1}^{\dagger}\phi_{2}-\phi_{2}^{\dagger}\phi_{1})\nonumber \\
	&& 
	+\lambda_{44}(F_3^{\dagger}F_3){\rm Tr}(\Delta^{\dagger}\Delta)+\lambda_{45}(F_3^{\dagger}F_3)(X^{\dagger}X)+\lambda_{46}(F_3^{\dagger}F_3)({X^{\prime}}^{\dagger}X^{\prime})++\lambda_{47}(F_4^{\dagger}F_4)(\phi_{1}^{\dagger}\phi_{1})\nonumber \\
	&& +\lambda_{48}(F_4^{\dagger}F_4)(\phi_{2}^{\dagger}\phi_{2})+\lambda_{49}(F_4^{\dagger}F_4)(\phi_{1}^{\dagger}\phi_{2}-\phi_{2}^{\dagger}\phi_{1})+\lambda_{50}(F_4^{\dagger}F_4){\rm Tr}(\Delta^{\dagger}\Delta)\nonumber \\
	&& 
	+\lambda_{51}(F_4^{\dagger}F_4)(X^{\dagger}X)+\lambda_{52}(F_4^{\dagger}F_4)({X^{\prime}}^{\dagger}X^{\prime})+\lambda_{53}(F_1^{\dagger}F_1)(F_2^{\dagger}F_2)+\lambda_{54}(F_1^{\dagger}F_1)(F_3^{\dagger}F_3)\nonumber \\
	&& +\lambda_{55}(F_1^{\dagger}F_1)(F_4^{\dagger}F_4)+\lambda_{56}(F_2^{\dagger}F_2)(F_3^{\dagger}F_3)+\lambda_{57}(F_2^{\dagger}F_2)(F_4^{\dagger}F_4) \nonumber \\
	&& +i m_{12}^2 (\phi_1^{\dagger}\phi_2-\phi_2^{\dagger}\phi_1) +\kappa_1 \tilde{\phi}_1^{T} i\sigma_2\Delta\phi_{1}+\kappa_2 \tilde{\phi}_2^T i\sigma_2\Delta\phi_2+ \kappa_3 (\tilde{\phi}_1^T i\sigma_2\Delta\phi_{2}+h.c.)  .
	\end{eqnarray}
	The VEV of all the scalar fields can be summarized as 
	\begin{eqnarray}
	\langle \phi_{1,2} \rangle =\frac{v}{\sqrt{2}}, \hspace{3mm} \langle \Delta \rangle =v_{\Delta}, \hspace{3mm} \langle X,X^{\prime} \rangle =v_{x,x^{\prime}}, \hspace{3mm} \langle F_i \rangle =v_{f_i}, \hspace{3mm}i=1,2,3,4.
	\end{eqnarray}
	Previously to explain the quark mass and mixing, we have taken $v_1=v_2=v=246/2=123$ GeV and $v_{x,x^{\prime}}=167$ GeV. Neutrino mass and mixing have been demonstrated with $v_{\Delta}=1$ eV, and UV completion has been done with the flavon fields having the breaking scale of 1 TeV, so $v_{f_i}\approx 1$ TeV. Now, one needs to adjust the unknown parameters so that the VEV of the correct order of magnitude can be obtained, which can be easily done as the potential consists of many unknown parameters. We can see that by taking all the $\lambda$ parameters to be $\mathcal{O}(1)$, we obtain the correct order of VEVs.
	\section{Model for CP violation}\label{sec:cp violation model}
	In section \ref {sec:neutrino mixing}, we build a model based on $CP\times Z_3$ where the neutrino mass matrix is real due to $CP$ symmetry of our model. However, the neutrino oscillation data give a hint of $CP$ violation which suggests that the $\delta_{\rm{CP}}$ should have some non-trivial value. The above model could not explain $CP$ violation as the $CP$ symmetry is exact in the neutrino sector. To explain $CP$ violation in the neutrino sector, our first task is to break the $CP$ symmetry spontaneously so that the Dirac CP-violating phase $\delta_{\rm{CP}}$ can take values other than $0$ or $\pi$. We know $CP$ symmetry implies $X\rightarrow X^*$, the exactness of $CP$ symmetry demands that the VEV of $X$ must be real. However, taking the VEV of $X$ to be complex breaks the $CP$ symmetry of our model spontaneously, and as a result, the complex phase of our model will be the source of $CP$ violation. So, let's start by writing the proposed Lagrangian, which is slightly different from Eq. (\ref{neutrino lagrangian}), which is allowed by the symmetries and particle contents of the model.
	\begin{eqnarray}\label{eq:CP breaking neutrino Lag}
	&&	\mathcal{L}_Y^{\prime} =\Big(\frac{X^{\prime}}{M}\Big)^3 Y_{ee}^{\nu} \bar{D}^c_{e L} i \sigma_2 \Delta D_{e L} + \Big(\frac{X}{M}\Big)^3Y_{e\mu}^{\nu} \bar{D}^c_{e L} i \sigma_2 \Delta D_{\mu L} +  \Big(\frac{X^{\dagger}}{M}\Big)^3 Y_{e \tau}^{\nu} \bar{D}^c_{e L} i \sigma_2 \Delta D_{\tau L}  \nonumber \\
	&& \Big(\frac{X^{\prime}}{M}\Big)^2 [Y_{\mu \mu}^{\nu} \bar{D}^c_{\mu L} i \sigma_2 \Delta D_{\mu L} +Y_{\tau \tau}^{\nu} \bar{D}^c_{\tau L} i \sigma_2 \Delta D_{\tau L} + Y_{\mu \tau}^{\nu} \bar{D}^c_{\mu L} i \sigma_2 \Delta D_{\tau L}] + h.c..
	\end{eqnarray}
	The phase of $X$ field now $\langle X \rangle =v_X e^{i\theta}$, $v_X$ is real. Other VEVs are same, as discussed in Section \ref{sec:neutrino mixing}. Then, the neutrino mass matrix takes the form
	\begin{eqnarray}\label{eq: new neutrino mass matrix}
	M_{\nu}=\begin{pmatrix}
	Y_{ee}^{\nu} \delta^3 &  Y_{e\mu}^{\nu}  \epsilon^3 e^{3i\theta} & Y_{e \tau}^{\nu} \epsilon^3 e^{-3i\theta} \\
	Y_{e\mu}^{\nu}\epsilon^3 e^{3i\theta} & Y_{\mu \mu}^{\nu}\delta^2 & Y_{\mu \tau}^{\nu}\delta^2 \\
	Y_{e\tau}^{\nu}\epsilon^3 e^{-3i\theta} & Y_{\mu \tau}^{\nu}\delta^2 & Y_{\tau \tau}^{\nu}\delta^2
	\end{pmatrix}v_{\Delta}.
	\end{eqnarray}
	Here also, all the couplings $Y^{\nu}$ are real due to $CP$ symmetry of the model. $\epsilon$ and $\delta$ are defined in section \ref{sec:neutrino mixing}.
	
	To diagonalize, we perform similar calculations as we have done before. We put $s_{12}^2=\frac{1}{3}$ and $s_{23}^2=\frac{1}{2}$ but here we choose the value of $\delta_{\rm{CP}}$ as $\delta_{\rm{CP}}=\frac{3\pi}{2}$ unlike the previous choice of $\delta_{\rm{CP}}(\pi)$. This choice of $\delta_{\rm{CP}}$ is called the maximal $CP$ violation. Now, this is a choice though but later we commented that the only allowed values of $\delta_{\rm{CP}}$ is $\frac{3\pi}{2}$. So, the unitary matrix that diagonalized the mass matrix of Eq. (\ref{eq: new neutrino mass matrix}) will take the form
	\begin{eqnarray}
	V_L^{\nu}=\begin{pmatrix}
	\sqrt{\frac{2}{3}}c_{13} & \frac{c_{13}}{\sqrt{3}} & i s_{13} \\
	-\frac{1}{\sqrt{6}}+i\frac{s_{13}}{\sqrt{3}} & \frac{1}{\sqrt{3}}+ i\frac{s_{13}}{\sqrt{6}} & \frac{c_{13}}{\sqrt{2}} \\
	\frac{1}{\sqrt{6}}+i\frac{s_{13}}{\sqrt{3}} & -\frac{1}{\sqrt{3}}+ i\frac{s_{13}}{\sqrt{6}} & \frac{c_{13}}{\sqrt{2}}
	\end{pmatrix}.
	\end{eqnarray}
	Neutrino mass diagonalization formula, as given in Eq. (\ref{inverted diagonalization formula}), and use the diagonalization procedure as prescribed in Section \ref{subsec:diagonalization}, we arrive at
	\begin{eqnarray}
	\frac{1}{m_a}M_{\nu}=\frac{1}{m_a}\begin{pmatrix}
	p & q & r\\
	q & s & b \\
	r & b & s
	\end{pmatrix},
	\end{eqnarray}
	where 
	\begin{eqnarray}
	{\rm{NO}} &:& p=\frac{1}{3}(2 m_1+m_2),\hspace{3mm}q=\frac{1}{6}\{-3\sqrt{2}im_3 s_{13}-2 m_1+2 m_2\},\nonumber \\
	&&	r=\frac{1}{6}\{-3\sqrt{2}im_3 s_{13}+2 m_1-2 m_2\},\hspace{3mm} s=\frac{1}{6}\{3 m_3+2 m_2+m_1\},\nonumber \\
	&& b=\frac{1}{6}(3 m_3-2m_2-m_1).\nonumber \\
	{\rm{IO}} &:& p=\frac{1}{3}(2 m_1+m_2),\hspace{3mm} q=\frac{1}{6} \{(2 m_1+2 m_2)+i(2\sqrt{2} m_1+\sqrt{2}m_2)s_{13}\},\nonumber \\
	&& r=\frac{1}{6} \{(2 m_1-2 m_2)+i(2\sqrt{2} m_1+\sqrt{2}m_2)s_{13}\},\hspace{3mm}s=\frac{1}{6}(3 m_3+2 m_2+m_1),\nonumber \\
	&& b=\frac{1}{6}(3 m_3-2m_2-m_1).
	\end{eqnarray}
	By looking at the above matrix elements and then comparing them with the previous matrix elements of Eq. (\ref {eq:expansion_mnu/ma}), only $q$ and $r$ are found to be different.
	
	Our next task is to find the numerical values of all the couplings $Y^{\nu}$ and $\theta$ by satisfying the neutrino oscillation data of Eq. (\ref{neutrino oscc data}). Again, we also follow the same prescription followed in section \ref{subsec:numerical analysis}. We just discussed that the matrix elements $p$, $s$ and $b$ are same as in Eq. (\ref{eq:expansion_mnu/ma}) as a result the result of Table. (\ref{tab:table1}) will be the same for this case. Only change will occur in $Y^{\nu}_{e\mu}$ and $Y^{\nu}_{e\tau}$. In addition to this, one new parameter $\theta$ is included, has to be found out as well. Before giving the numerical values, one comment: here we have obtained the values of $Y_{e\mu}$, $Y_{e\tau}$ and $\theta$ for the NO of neutrino masses because the whole calculation of our model of section \ref{sec:neutrino mixing} discarded IO of neutrino masses. Using the best-fit value of $\sin\theta_{13}$, solar and atmospheric mass squared differences for normal ordering of neutrino masses, we get
	\begin{eqnarray}\label{eq:new numerical}
	&&(Y^{\nu}_{e\mu},\hspace{2mm} Y^{\nu}_{e\tau},\hspace{2mm}\theta)\approx (1.188,\hspace{2mm}1.188,\hspace{2mm}25.86^{\circ}),\hspace{5mm} (m_1\neq 0)\nonumber \\
	&& (Y^{\nu}_{e\mu},\hspace{2mm} Y^{\nu}_{e\tau},\hspace{2mm}\theta)\approx (1.451,\hspace{2mm}1.451,\hspace{2mm}17.69^{\circ}). \hspace{5mm} (m_1= 0)
	\end{eqnarray}
	Here the results suggest that our model can explain maximal $CP$ violation provided $Y^{\nu}_{e\mu}=Y^{\nu}_{e\tau}$. Now, with this condition, the neutrino mass matrix of Eq. (\ref{eq: new neutrino mass matrix}) looks like $\mu-\tau$ symmetric mass matrix \cite{Xing:2015fdg} the prediction of which is that only maximal $CP$ violation is possible. So, the noticing point is that our model can have only maximal $CP$ violating phase. If we vary all the neutrino oscillation observables according to Eq. (\ref{neutrino oscc data}), the region for the allowed range of $Y_{e\mu}$, $Y_{e\tau}$ are shown in Fig. \ref{fig:new-Y^emu, Y^etau}.
	\begin{figure}[h!]
		\begin{center}
			\includegraphics[height=2in,width=3in]{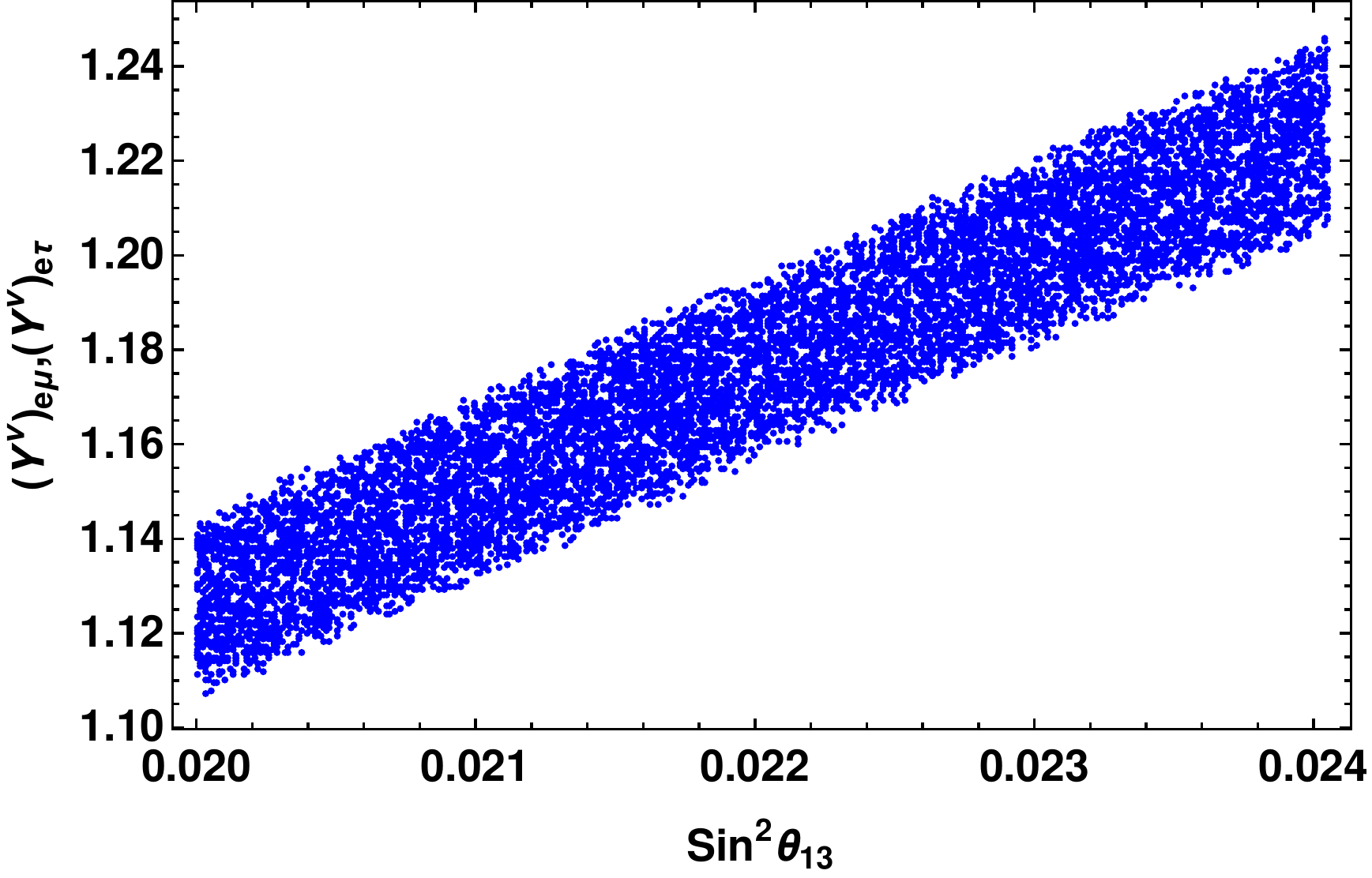}
			\includegraphics[height=2in,width=3in]{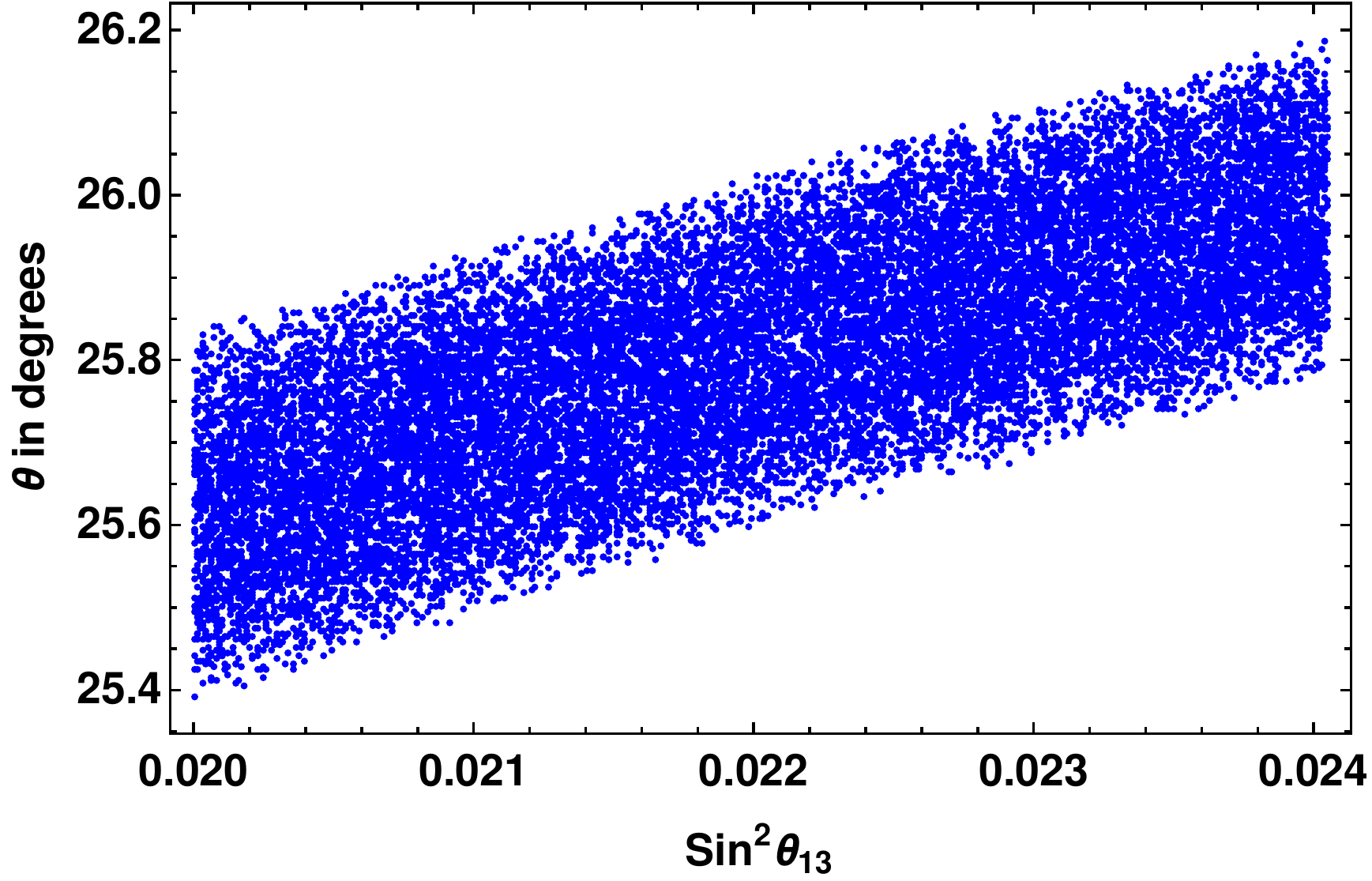}
		\end{center}
		\caption{Two couplings $Y^{\nu}_{e\mu}$ and $Y^{\nu}_{e\mu}$(left), and $\theta$(right) are plotted with the $3\sigma$ allowed range of $\sin^2\theta_{13}$. The blue regions are allowed.}
		\label{fig:new-Y^emu, Y^etau}
	\end{figure}
	
	Now, our last task is to show how the Lagrangian can be motivated from the UV completion mechanism as done for the previous case in section \ref{subsec:UV completion} in full detail. UV completion mechanism is also needed here because there are some tree-level or higher-dimensional terms allowed by the symmetries of our model other than given in Eq. (\ref{eq:CP breaking neutrino Lag}). Here, we give only the prescription for doing the UV completion as the procedure will be the same as in section \ref{subsec:UV completion}. There $U(1)_X$, $U(1)_F$ along with some flavons and vector-like leptons were introduced, and then the renormalization Lagrangian was written, which includes these heavy fields. After integrating out these heavy fields, the Lagrangian of Eq. (\ref{eq:CP breaking neutrino Lag}) will be generated. Here the only difference is that we do not introduce the $U(1)_X$ symmetry because $U(1)_X$ symmetry will make the VEV of $X$ field real, and then CP violation cannot be explained. So, with the exclusion of $U(1)_X$ symmetry, now, the full scalar potential will change as well, but this will not change our conclusions; essential terms like $X^3$ or $X^2 X^{\prime}$ and many more in the scalar potential are allowed where the phase $\theta$ will appear explicitly. Then, minimizing the scalar potential w.r.t $\theta$ can determine the value of $\theta$ of Eq. (\ref{eq:new numerical}) by tuning the parameters of the scalar potential. 
	\section{Conclusion}\label{sec:conclusion}
	In this work, we commented on the quark mass and mixing patterns by following the mass matrix texture of Ref. \cite{Babu:1999me}. Within the same framework, we built a model using non-renormalizable operators based on the type II seesaw mechanism to explain neutrino mass and mixing. We fit the neutrino oscillation data, and the model prefers NO of neutrino masses. In this model, the exact $CP$ symmetry forbids $CP$ violation. However, we demonstrated that the maximal $CP$ violation could be incorporated into our model by slightly modifying the Lagrangian within the same model. Then we have shown that higher-dimensional terms can be motivated by the UV completion of our model. Within this model, we calculated the branching fraction of LFV decays and found the lower bound of the mass of the triplet scalar. At the end, the full scalar potential of the model is discussed with some phenomenological implications. 
	\section*{Acknowledgement}
	I acknowledge Dr. Raghavendra Srikanth Hundi for his helpful comments. I am also thankful to Satyabrata Mahapatra for some valuable discussions and helpful comments while writing the draft.
	
\end{document}